\documentclass[aps,twocolumn,prd,reprint,superscriptaddress,showpacs,preprintnumbers,nofootinbib,10pt,floatfix]{revtex4-1}
\RequirePackage[l2tabu, orthodox]{nag}

\usepackage[utf8x]{inputenc}
\usepackage{microtype} 
\usepackage{amsmath}
\usepackage{amstext, amssymb, amsthm, amsfonts, slashed}
\usepackage{mathtools}
\usepackage{mhchem}
\usepackage{graphicx}
\usepackage{{booktabs}}

\usepackage{dcolumn}
\usepackage{bm}
\usepackage{dsfont}
\usepackage[usenames,dvipsnames]{xcolor}
\usepackage[normalem]{ulem}

\usepackage{url}
\usepackage[colorlinks=true,urlcolor=blue,linkcolor=blue,citecolor=blue,hypertexnames]{hyperref}
\usepackage[noabbrev]{cleveref}

\graphicspath{ {./graphics/} }

\DeclareSymbolFont{symbolsC}{U}{pxsyc}{m}{n}
\DeclareMathSymbol{\colonequ}{\mathrel}{symbolsC}{"42}
\DeclareMathOperator{\Tr}{Tr}

\newcommand{\vect}[1]{\bm{#1}}
\newcommand{\imaginaryi}{i}
\newcommand{\al}[1]{\begin{align}#1\end{align}}
\newcommand{\bvec}[1]{\mbox{\boldmath $#1$}}
\newcommand{\p}{\partial}
\newcommand{\nn}{\nonumber\\} 

\newcommand{\fn}[1]{\!\left(#1\right)}

\begin{document}
\title{Partial bosonisation for the two-dimensional Hubbard model: How well does it work?}
\author{Tobias \surname{Denz}}
\affiliation{Institut f{\" u}r Theoretische Physik, Universit{\" a}t Heidelberg, Philosophenweg 16, 69120 Heidelberg, Germany}
\author{Mario \surname{Mitter}}
\affiliation{Institut f{\" u}r Theoretische Physik, Universit{\" a}t Heidelberg, Philosophenweg 16, 69120 Heidelberg, Germany}
\author{Jan M. \surname{Pawlowski}}
\affiliation{Institut f{\" u}r Theoretische Physik, Universit{\" a}t Heidelberg, Philosophenweg 16, 69120 Heidelberg, Germany}
\affiliation{ExtreMe Matter Institute EMMI, GSI Helmholtzzentrum f{\" u}r Schwerionenforschung mbH, Planckstr. 1, 64291 Darmstadt, Germany}
\author{Christof \surname{Wetterich}}
\affiliation{Institut f{\" u}r Theoretische Physik, Universit{\" a}t Heidelberg, Philosophenweg 16, 69120 Heidelberg, Germany}
\author{Masatoshi \surname{Yamada}}
\affiliation{Institut f{\" u}r Theoretische Physik, Universit{\" a}t Heidelberg, Philosophenweg 16, 69120 Heidelberg, Germany}

\begin{abstract}
Partial bosonisation of the two-dimensional Hubbard model focuses the functional renormalisation flow on channels in which interactions become strong and local order sets in.
We compare the momentum structure of the four-fermion vertex, obtained on the basis of a patching approximation, to an effective bosonic description.
For parameters in the antiferromagnetic phase near the onset of local antiferromagnetic order, the interaction of the electrons is indeed well described by the exchange of collective bosonic degrees of freedom.
The residual four-fermion vertex after the subtraction of the bosonic exchange contribution is small.
We propose that similar partial bosonisation techniques can improve the accuracy of renormalisation flow studies also for the case of competing order.
\end{abstract}
\maketitle

\section{Introduction}
\label{sec: introduction}
Cuprates exhibiting high-$T_\text{C}$ superconductivity~\cite{1986ZPhyB..64..189B,Wu:1987te} feature effectively two-dimensional \ce{CuO2} layers.
For this reason, Anderson~\cite{Anderson:1987gf}, as well as Zhang and Rice~\cite{Zhang:1988yua}, proposed using the two-dimensional Hubbard model on a square lattice~\cite{1963RSPSA.276..238H, 1963PhRvL..10..159G, 1963PThPh..30..275K} to describe the phase diagram of these superconductors.
Despite the Hubbard model's conceptual simplicity, only a few features are known rigorously for special cases~\cite{1998JPCM...10.4353T}. 
A quantitative understanding of the macroscopic properties of this simple microscopic model has remained a great challenge for theoretical many-body physics.
In particular, it is expected that competing order effects of the strongly correlated fermions lead to a particularly rich phase structure. 

Functional renormalisation group (FRG) techniques have already made important contributions~\cite{1987EL......4..609S,lederer:jpa-00210598,Dzyaloshinskii1987JETP,1996JPhy1...6..119D,%
PhysRevLett.81.3195,1997ZPhyB.103..339Z,1998EL.....44..235Z,Halboth:2000zza,Halboth:2000zz,PhysRevB.61.13609,2001EPJB...21...81H,Salmhofer:2001tr,PhysRevB.63.035109,PhysRevB.64.184516,PhysRevB.72.205128,%
2006PSSBR.243...46M,PhysRevB.75.075110,PhysRevB.79.094526,%
Baier:2000yc,Baier:2001wq,Baier:2003ex,Baier:2003fw,Krahl:2006tx,Friederich:2010zn,Friederich:2010hr,%
Krahl:2009zp,PhysRevB.85.075121,PhysRevB.86.245122,PhysRevB.86.235140,PhysRevB.89.035126,PhysRevB.89.121116,PhysRevLett.116.096402,PhysRevB.86.085113,PhysRevB.95.085143,PhysRevB.97.235110,PhysRevB.98.235131,PhysRevB.98.075143,10.21468/SciPostPhys.6.1.009,PhysRevB.99.104501,PhysRevB.100.014504,2019NuPhB.941..868S} to a qualitative and sometimes quantitative understanding of the two-dimensional Hubbard model on a square lattice.
For a computation of the phase structure of the Hubbard model it is crucial to precisely take into account fluctuations beyond the mean-field approximation.
Because of its non perturbative and versatile character, it is expected that the FRG is well suited to capture the relevant dynamics.
In parallel, numerical simulations based on the Monte Carlo method have provided first-principles computations for such strongly correlated systems~\cite{PhysRevB.31.4403,PhysRevB.40.506,PhysRevB.39.839,PhysRevB.41.2313,2002AdPhy..51.1587B,PhysRevB.48.3380,PhysRevLett.75.1344,PhysRevLett.79.1122,PhysRevB.56.5597}.
It is difficult, however, to use them to investigate the full phase structure of the Hubbard model due to the sign problem in systems with a finite chemical potential.
Furthermore, diagrammatic resummation techniques as dynamical mean-field theory (DMFT) have provided insight into important structures of the Hubbard model~\cite{PhysRevB.100.155149,PhysRevB.100.245147,PhysRevB.101.075109}.
They can be combined with FRG, called DMF$^2$RG~\cite{PhysRevLett.112.196402}.
(For a compilation, see \cite{RevModPhys.68.13} and references therein.)

The Hubbard model is defined as a purely fermionic model which describes the hopping and scattering of electrons on a lattice.
It provides a natural description of the relevant microscopic electron dynamics in solids.
On the other hand, the macroscopic properties, reflected in various possible order parameters, and long-range dynamics, are most easily described by collective bosonic excitations such as antiferromagnetic spin waves or $d$-wave electron pairs.
Order parameters can be interpreted as the expectation values of these composite bosonic fields.

Within the FRG, two main approaches have been followed.
The first concentrates on the flow of the four-fermion interaction in a purely fermionic setting.
On microscopic scales, one starts with a point-like interaction.
As more and more fluctuations are included, the renormalisation flow generates a complicated momentum dependence of this vertex.
Following numerically the flow of the momentum-dependent interaction, one observes strong enhancements, and even divergences, in certain momentum channels.
This is related to the possible onset of condensation phenomena in the corresponding channels.
First set up for a description of collective mesons in a purely fermionic quark model~\cite{Ellwanger:1994wy}, this method has found important developments for strongly correlated electrons.
In particular, the $N$-patching method~\cite{1997ZPhyB.103..339Z} has allowed for a better resolution of the momentum dependence of the four-fermion coupling and provided many insights for the Hubbard model~\cite{1997ZPhyB.103..339Z,1998EL.....44..235Z,Halboth:2000zza,Halboth:2000zz}, and other strongly correlated electron systems like, e.g., pnictides~\cite{2009EL.....8537005W,PhysRevB.81.184512,PhysRevLett.102.047005} or graphene~\cite{PhysRevB.86.155415,PhysRevLett.109.126402,PhysRevB.85.235408,PhysRevB.99.195120}.

The second main approach has employed some form of bosonisation~\cite{Aoki:1999dw,Baier:2000yc,Gies:2001nw,Gies:2002hq,Baier:2001wq,Baier:2003ex,Baier:2003fw,Pawlowski:2005xe,Krahl:2006tx,Friederich:2010hr,Friederich:2010zn,PhysRevB.79.094526,Floerchinger:2009uf,Braun:2014ata,Fu:2019hdw,Mitter:2014wpa,Cyrol:2017ewj,2017arXiv171204297W}.
By means of a Hubbard-Stratonovich transformation~\cite{1957SPhD....2..416S,Hubbard:1959ub}, the fermionic system with a point-like interaction can be rewritten in terms of one or several ``collective'' bosons coupled to fermions by Yukawa couplings.
For the microscopic point-like interaction one can employ a standard form of the momentum dependence for the fermion-boson coupling, reflecting the properties of the corresponding channel.
At this stage, the boson propagator is simply a constant.
As the flow proceeds to smaller momentum scales, the inclusion of fluctuation effects generates a momentum dependence of the boson propagators.
Furthermore, the bosonic part of the effective action does not remain quadratic in the boson fields. 
A non-trivial effective potential develops for these fields.
A minimum of the potential at non-zero field value indicates spontaneous symmetry breaking, with the corresponding order parameter given by the value of the bosonic field at the minimum of the potential.
The flow can be followed into the ordered phase, and the divergences in the four-fermion vertex at the onset of local order can be avoided.
This has allowed to establish antiferromagnetic and $d$-wave superconducting order in corresponding regions of parameter space~\cite{2006PSSBR.243...46M,PhysRevB.75.075110,Baier:2000yc,Baier:2001wq,Baier:2003ex,Baier:2003fw,Krahl:2006tx,PhysRevB.79.094526,Friederich:2010hr,Friederich:2010zn}, and the existence of competing order~\cite{Friederich:2010hr}.

An important aspect of the bosonised approach is the possibility of ``dynamic bosonisation"~\cite{Gies:2001nw, Gies:2002hq,
  Pawlowski:2005xe, Floerchinger:2009uf, Braun:2014ata, Fu:2019hdw}.
Fluctuations generate four-fermion interactions even if the original point-like four-fermion interaction has been bosonised via a Hubbard-Stratonovich transformation and replaced by boson-fermion interactions.
Flowing bosonisation allows to (partially) absorb the newly generated four-fermion vertex in the flow of the propagator and the Yukawa couplings of the interacting system of fermions and bosons.
Using dynamical bosonisation, there is no need to start with a Hubbard-Stratonovich transformation.
One may keep the fermionic interaction, and only attribute dominant fluctuation effects to a suitable boson exchange~\cite{Friederich:2010zn,Friederich:2010hr}.

 Both approaches have their advantages and disadvantages.
 The purely fermionic description allows for a detailed and unbiased resolution of the momentum dependence of the four-fermion vertex.
 No {\it a priori} guess on relevant channels is necessary.
 The disadvantage is a rather indirect description of spontaneous symmetry breaking.
 Typically, the quartic four-fermion vertex diverges at the onset of local spontaneous symmetry breaking, and the flow cannot be continued reliably into the ordered phase.

In contrast, a bosonic description is well suited for a description of
spontaneous symmetry breaking.  The effective potential develops terms
that are quartic in the bosonic fields.  Such a potential can account
for spontaneous symmetry breaking whenever the term quadratic in the
boson fields, the boson ``mass term,''  becomes negative.  The
vanishing of the mass term at the onset of spontaneous symmetry
breaking produces the divergence of the four-fermion vertex in the
corresponding purely fermionic description.  Quartic boson
interactions, which stabilise the effective potential, correspond to
vertices with eight fermions in the purely fermionic language, which
are hard to incorporate in the purely fermionic flow. 

The disadvantage of bosonic methods is the necessity of guessing the
relevant channels corresponding to collective bosons.
In most applications, the remnant four-fermion interaction is dropped. Since the
relevant channels may differ for different parameters of the
macroscopic model, and several channels may be equally important, the
guess introduces a certain bias in the description.  For example, in
mean-field theory results depend strongly on the choice of the
channels in the Hubbard-Stratonovich transformation~\cite{Baier:2001wq,Baier:2003ex,Baier:2003fw}.  This
dependence is greatly reduced once fluctuations of the collective
bosons are included~\cite{Jaeckel:2002rm}, but a residual bias remains
for a given truncation (see, e.g.,~\cite{Pawlowski:2015mlf}).
Furthermore, the exchange of collective bosons restricts the possible
momentum dependence of the effective four-fermion interaction
generated by the boson exchange.
In a more physical sense, one has to guess what kind of order parameter could be realised in the vacuum state of the system and introduce the corresponding collective (auxiliary) fields.
In other words, in this method, one has to guess in advance which symmetries are likely to be broken by the dynamics of electrons.
\begin{table}
\begin{center}
\begin{tabular}{ l c c c c c }
\toprule
  \makebox[1.1cm]{} &   \makebox[1.1cm]{SSB}  &   \makebox[1.1cm]{FM}  & \makebox[1.1cm]{FL}  &   \makebox[1.1cm]{BF}  \\
\midrule
PF & 			 & $\checkmark$ & & $\checkmark$  \\[2ex]
HS &  $\checkmark$ &  & \\[2ex]
DB &   $\checkmark$ & & $\checkmark$ & \\[2ex]
DBF & $\checkmark$ & $\checkmark$ & $\checkmark$ & $\checkmark$  \\
\bottomrule
\end{tabular}
\caption{
\label{Table: summery of methods}
Comparison between different methods.
The abbreviations are given as follows. (PF): pure fermionic theory; (HS): bosonisation via the Hubbard-Stratonovich transformation without residual four-fermion interaction; (DB): dynamical bosonisation without residual four-fermion interaction; (DBF): dynamical bosonisation with residual four-fermion interaction; (SSB): naturally access to spontaneous symmetry breaking with order parameter; (FM): full momentum dependence of complete four-fermion vertex; (FL): fluctuations on all scales included for leading channels; (BF): free of bias of channel selection.
The check mark ($\checkmark$) indicates that a method can naturally address to an issue. 
}
\end{center}
\end{table}

This has led to approaches that combine the introduction of resonant
channel with the fermionic description: 
Within a purely fermionic description an organisation of the
momentum-dependent four-fermion interaction in terms of important
collective channels has been advocated~\cite{Salmhofer:2001tr}. In QCD
a Fiertz-complete basis of the four-fermion interaction has been taken
into account, while treating the resonant scalar-pseudoscalar channel
with dynamical bosonisation, \cite{Mitter:2014wpa,Cyrol:2017ewj}. Such
a Fiertz complete approach with full momentum dependence requires
sophisticated numerical tools, and the numerics in this paper
is based on the numerical tools developed in \cite{Mitter:2014wpa,Cyrol:2017ewj} for relativistic systems. The
setup allows in particular to check whether or not additional channels
are resonant. In QCD in the vacuum this is not the case and no
competing order effects occur. For the Hubbard model, the problem of competing order
parameters has been studied in Refs.\,\cite{2006PSSBR.243...46M,PhysRevB.75.075110} within an effective
mean-field theory for the momentum modes that have not yet been included at the scale where the four-fermion vertex diverges. For investigations of competing order on the bosonised side the idea is to use only a partial bosonisation, while keeping, in addition, a four-fermion interaction.  This approach has been followed
in~\cite{Friederich:2010zn,Friederich:2010hr}, where a
momentum-independent four-fermion interaction is kept in addition to
the one generated by boson exchange.
In \cref{Table: summery of methods} we summarise comparisons between different methods

In this paper we propose to combine partial bosonisation with a
momentum-dependent residual four-fermion interaction.
The flow of the four-fermion action can be followed by use of an $N$-patching scheme.
We suggest a truncation that involves a combined system of fermions and bosons, for which the bosonic fields represent fermion bilinears or more complicated collective fields.
We propose to keep a momentum-dependent four-fermion vertex in addition
to the flow of boson propagators, Yukawa interactions, and an effective
potential for the bosons.
Partial bosonisation can be employed to convert
the four-fermion vertex into the exchange of collective bosons as much
as possible.  This approach preserves all the advantages of a bosonic
description.  In addition, the momentum resolution of the total
four-fermion interaction is no longer limited since it appears in the
residual four-fermion interaction, which remains even after dynamical
bosonisation.  There remains a residual bias in the selection of the
channels in which boson fields are introduced.  Only for these
channels, higher-order terms in the bosonic fields can represent
corresponding eighth-order fermionic interactions.  The consequences
of this bias are much reduced, however, since the momentum dependence
of the residual four-fermion interaction can still ``detect" important
channels that may have been omitted by the guess for the bosonised
channels.

\begin{figure*}
  \includegraphics[width = \linewidth]{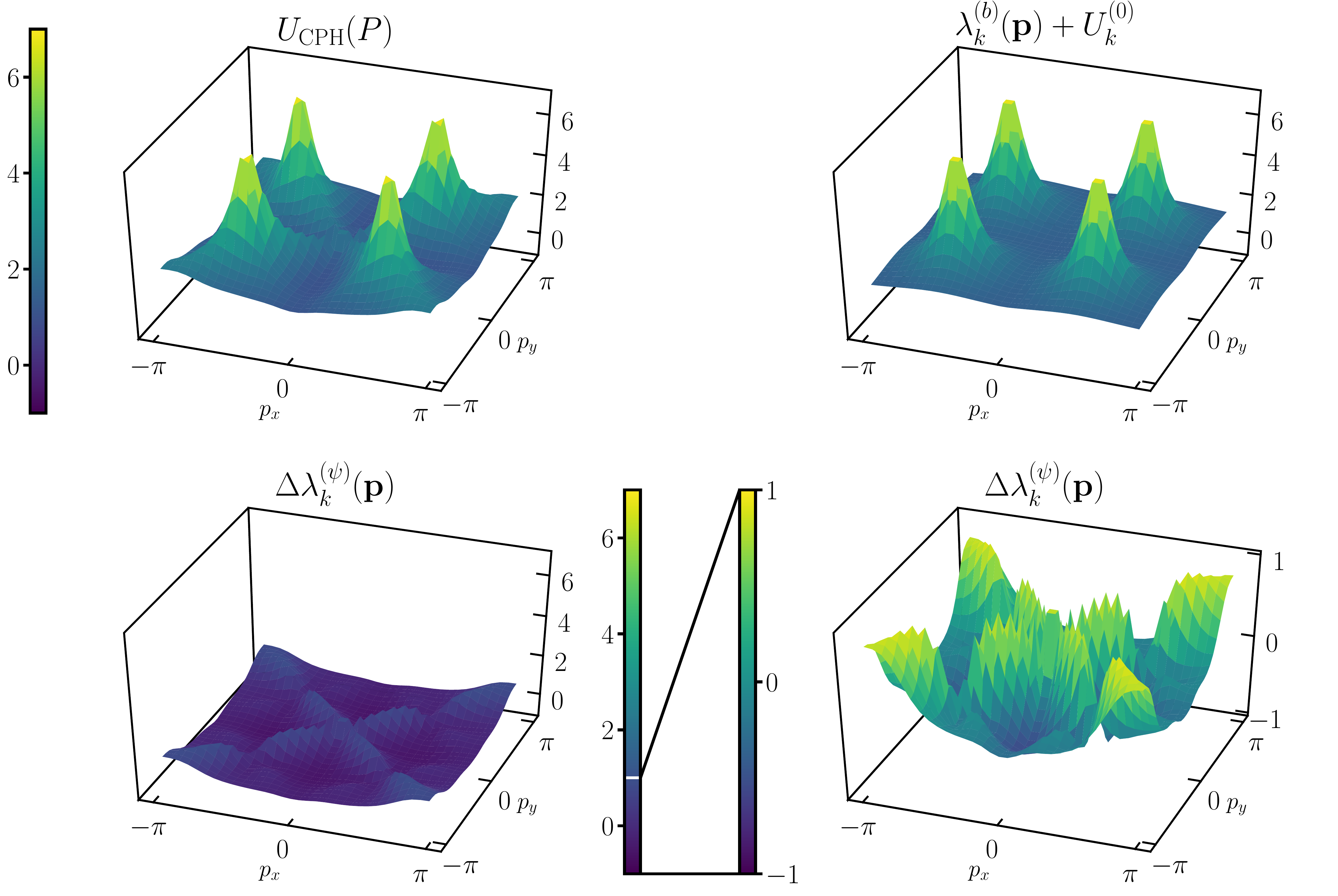}
  \caption[Four-fermi coupling in the IR]{ Momentum dependence of the four-fermion vertex. Parameters are $\mu=t'=0$, $U/t=3$, $T=0.1$, and the vertex is evaluated at $k_{\text{IR}} \approx 0.9 t$ ($T_k \approx 0.25 t$).
  We display $U_k(P,P,-P,-P)=U_\text{CPH}\fn{P}$, where at one-loop level only the crossed particle-hole channel (CPH) contributes, with $P=(0,p_x,p_y)$.
  The top left panel shows the numerical result, which is dominated by nesting peaks located at $(\pm\pi/2,\pm\pi/2)$.
  The top right panel shows a boson exchange mediated vertex of the form $\lambda_k^{(b)}(\vect{p})+U_k^{(0)} = 2(m_k^2+A_k[2\vect{p}-{\vect\pi}]^2)^{-1} + \lambda_k^{(0)}$, where $m_k^2$, $A_k$, and $\lambda_k^{(0)}$ are momentum-independent fitting parameters.
  The bottom panels show the momentum dependence of the residual four-fermion interaction, given by the difference between the two vertices on the top panels,  $\Delta\lambda_k^{(\psi)}\fn{\vect p}=U_\text{CPH}\fn{P}-(\lambda_k^{(b)}\fn{\vect p}+U_k^{(0)})$.  The bottom right panel is a magnified version of the one on the bottom left.
Differences between $U_\text{CPH}\fn{P}$ and $\lambda_k^{(b)}\fn{\vect p}+U_k^{(0)}$ are overall about one order of magnitude smaller than the peaks.
}
  \label{fig:ff_coupling_IR_main_figure}
\end{figure*}

A crucial ingredient for the proposed scheme of partial bosonisation
is the relative size of the boson-mediated and the residual
four-fermion interaction.  This is the focus of the present paper.
Partial bosonisation is most efficient if the leading part of the full
effective four-fermion interaction can be expressed in terms of boson
exchanges, with a comparatively small residual four-fermion
interaction.  Here, we concentrate on the antiferromagnetic phase for
which the exchange of bosons associated to antiferromagnetic spin
waves is indeed found to be the dominant interaction.

The main result of this work is demonstrated in
\cref{fig:ff_coupling_IR_main_figure}.  The upper left subfigure shows
the momentum dependence of the four-fermion vertex in an appropriate
projection in the space of momenta of the fermions. 
It has been computed by solving the purely fermionic flow using the $N$-patching method down to a particular scale $k_\text{IR}$.
At this scale we investigate to which extent the four-fermion vertex can be accounted for by boson exchange.
The upper right plot displays the contribution of a boson
exchange, expressed via a generic boson propagator with three fitted
parameters.  The residual four-fermion vertex
after subtraction of the boson exchange contribution is presented in
the lower left.  It is substantially smaller than the boson exchange
contribution, and shows less structure, as shown in the magnified plot
in the lower right.  All quantities are evaluated at a momentum scale
$k_{\text{IR}}$ of the flow close to the onset of local order.  More
details on this are given in \cref{main result}.
This figure suggests that the four-fermion vertex is dominated by the exchange of a collective degree of freedom called antiferromagnetic boson.
An investigation of other momentum channels confirms this picture.
Combining three different momentum channels we find, in addition, indications for an $s$-wave pairing boson to play a role, being subleading as compared to the antiferromagnetic boson.
All three momentum channels are simultaneously well described by the exchange of antiferromagnetic and $s$-wave pairing bosons, together with a constant four-fermion interaction. The momentum dependence of the residual four-fermion interaction is found to be small.

This paper is organised as follows: In \cref{main result} we present
our main findings by taking as a specific example a region in
parameter space where antiferromagnetic order is expected.
\Cref{sec:hubbard_model} specifies our notation for the Hubbard model
and \cref{sec:frg} recapitulates the relevant features of functional
renormalisation. \Cref{Sec: solving pure fermionic equation} describes the fermionic flow, and \cref{sect: Partial bosonisation}
discusses partial bosonisation.  Finally, we summarise our findings in
\cref{sec:conclusions}.

\section{Partial bosonisation of the four-fermion interaction}
\label{main result}
Partial bosonisation has been employed to describe transitions in the
effective degrees of freedom within the FRG.  The main idea is to
rewrite the ``four-fermion vertex" or ``two-particle vertex", or parts of it, in terms of exchanges of an antiferromagnetic boson:
\begin{align}
  \vcenter{\hbox{\includegraphics[width=0.87 \linewidth]{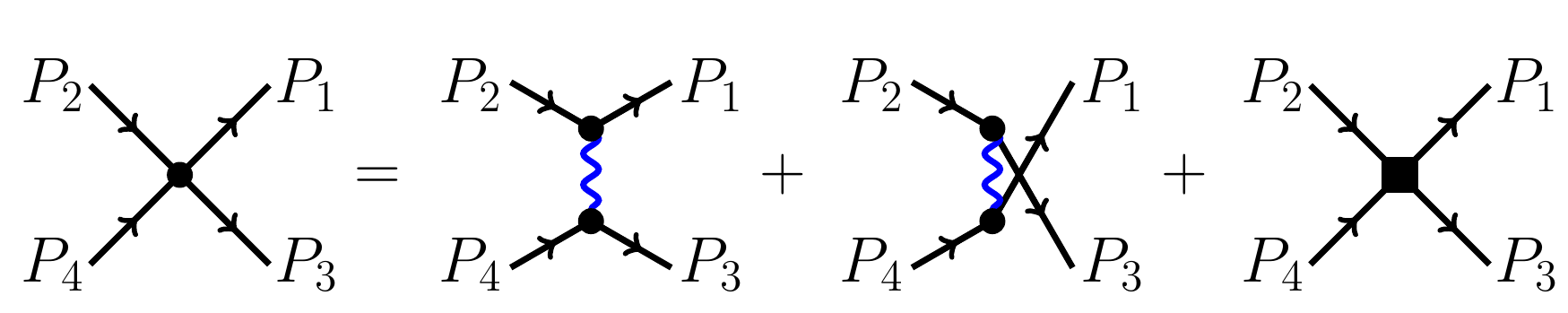}}} \;.
\label{eq: diagrammatic bosonisation process}
\end{align}
Here, the left-hand side is the four-fermion interaction given in the
pure fermionic system.
The first diagram on the right-hand side can be interpreted as an electron and a hole forming an ``antiferromagnetic boson", which then decays into an electron and a hole again.
The second diagram on the right-hand side accounts for electron-hole scattering by exchange of the antiferromagnetic boson.
Finally, the third diagram on the right-hand
side denotes the residual four-fermion vertex after the subtraction of
the boson exchange contribution from the full vertex. We can interpret this equation in both directions. We can either employ it to decompose a given vertex into a bosonic and a residual part.
Or, in the bosonic formulation, it accounts for the full effective
vertex in terms of contributions from boson exchange and an
``irreducible" part. Formally, it constructs the connected four-point
function from one-particle irreducible (1PI) parts. Note in this
context that in the purely fermionic language the total vertex is
1PI. There is some freedom in the choice of the split between boson
exchange and residual four-fermion interaction, which can be used to
optimise the procedure. Observable results, such as the gap in the
fermion propagator due to collective order, have to be independent of
the choice of the split. This can serve as a partial error estimate
for a given approximation.

Dynamical bosonisation provides for an exact formal
method~\cite{Gies:2001nw, Gies:2002hq, Pawlowski:2005xe,
  Floerchinger:2009uf, Braun:2014ata, Fu:2019hdw} of translating parts
of the four-fermion interaction to an effective action for bosons and
fermions.
It can always be implemented.
However, the crucial question is under which circumstances it is useful.  If one keeps the momentum dependence of the residual four-fermion vertex, the addition of bosonic degrees of freedom and their interactions first adds to the complexity of the problem.
Partial bosonisation will only be useful if the most relevant parts of the four-fermion interaction are captured by the boson exchange, while it seems superfluous if the residual interaction dominates.
The relative importance of the boson exchange is a quantitative
question that we investigate here by solving the flow of the
four-fermion interaction numerically, and by comparing the size of the boson exchange and the residual four-fermion interaction at the scale $k_\text{IR}$.

The central quantity for this problem is the four-fermion interaction
[see \cref{equ:action_bare} in \cref{sec:hubbard_model}], which we
parametrise by a momentum-dependent coupling
$U_k\fn{P_1,P_2,P_3,P_4}$.  Here, $P_i$, $i=1,...,4$, are the momenta
and frequencies of the fermions in the vertex, and the index $k$
indicates that we consider a scale-dependent coupling which flows from
a microscopic scale to a macroscopic scale as the effects of
fluctuations are successively incorporated.  The corresponding term in
the scale-dependent effective action (effective average action or coarse grained free energy)
$\Gamma_k$ reads
\begin{align} \label{equ:action_int}
  \Gamma_k^\text{int}&= \frac{1}{2}\sum_{P_i}
                       \delta(P_1-P_2+P_3-P_4) \,U_k(P_1,P_2,P_3,P_4) \nonumber \\[1ex]
                     &\qquad\times
                       \psi^\dag_\alpha(P_1)\psi^{}_\alpha(P_2) \,
                       \psi^{\dag}_\beta(P_3) \psi^{}_\beta(P_4) \;,
\end{align}
where $P_i=(\omega, p_x, p_y)_i$ is the short-hand notation
collecting the Matsubara frequency
$\omega$ and
a two-dimensional (2D) momentum $\bvec{p}=(p_x,p_y)$ for the electron field
$\psi_\alpha$, where the indices $\alpha$, $\beta$ denote
the electron spin.  More concrete setups and definitions for the FRG
and the Hubbard model are presented in the following sections.  The
initial value of $U_k$ at the microscopic scale $k=\Lambda$ is given by the Hubbard coupling $U$ for a pointlike interaction, which is a
momentum-independent constant
\begin{align}
U_\Lambda\fn{P_1,P_2,P_3,P_4}=U\,,
\end{align}
with $\Lambda$ of the order of the inverse lattice distance.  Using
the FRG equation with an $N$-patching scheme for the four-fermion
vertex, we follow the RG flow of $U_k(P_1,P_2,P_3,P_4)$ from a ultraviolet (UV)
scale $k=\Lambda$ to a certain infrared (IR) scale $k=k_\text{IR}$.  
We do not resolve the frequency dependence here, working at $\omega_i=0$.
For a general momentum configuration, $U_{k_\text{IR}}(P_1,P_2,P_3,P_4)$ is then
a non-trivial function of three independent two-dimensional momenta.

Partial bosonisation should capture the non-trivial momentum
dependence of $U_k(P_1,P_2,P_3,P_4)$ in terms of boson exchange
processes.  Let us add to the effective action a bosonic part
\begin{align}
  &\Gamma_k^a = \sum_{Q,j} \bigg\{ \frac{1}{2}\phi_j\fn{-Q} \left( \hat{G}^{(a)}_k(Q) \right)^{-1} \phi_j\fn{Q} \nn
  &
    -\sum_{P_1,P_2}\delta\fn{Q-P_1+P_2}\psi^{\dag}_\alpha\fn{P_1}(\sigma^j)_{\alpha\beta}
    \psi_\beta\fn{P_2}\phi_j\fn{Q}\bigg\}\,.
\end{align}
The three boson fields
$\phi_j\fn{Q}$ correspond to the Fourier modes of spin waves in the
spin one channel, with $\sigma^j$ the Pauli matrices. For the
particular momentum $Q=\Pi=(0,\pi,\pi)$, this is an antiferromagnetic
spin wave. The inverse propagator of a boson of this type is taken to be
\begin{align} \label{equ:propagator_boson_af}
\left( \hat{G}^{(a)}_k(Q) \right)^{-1}=m_k^2+A_k[Q-\Pi]^2\,,
\end{align}
with $[P]^2=p_x^2+p_y^2$ if $p_x,p_y\in [-\pi,\pi]$, and periodically
continued outside the Brillouin zone.  
The superscript $(a)$ denotes the``antiferromagnetic" type boson.
It is chosen to have its
minimum at $Q=\Pi$ if $A_k>0$. 
A non-zero expectation value $\langle\phi_j\fn{Q=\Pi}\rangle$ corresponds to antiferromagnetic order.
The Yukawa coupling of the bosons to the
electrons is set to one here.

The field equation for $\phi_j$ in the presence of fermion fields reads as
\begin{align}
\phi_j\fn{Q}=\sum_{P_1} \hat{G}^{(a)}_k(Q) \psi^{\dag}\fn{P_1}\sigma^j \psi\fn{P_1+Q}\,.
\label{eq: field equation of motion for phi}
\end{align}
This identifies $\phi_j$ with a fermion bilinear.
A Gaussian integration over the boson field amounts to insertion of the field equation \eqref{eq: field equation of motion for phi} into the effective action $\Gamma_k^a$, resulting in
\begin{align}
\Gamma_k^a = - \frac{1}{2} \sum_j \sum_{Q,P_1,P_3}& \hat{G}^{(a)}_k(Q) [\psi^{\dag}\fn{P_1}\sigma^j\psi\fn{P_1-Q}]\nn
&\times[\psi^{\dag}\fn{P_3}\sigma^j\psi\fn{P_3+Q}]\,.
\end{align}
With the identity
\begin{align}
\sum_j(\sigma^j)_{\alpha\beta}(\sigma^j)_{\gamma\delta}=2\delta_{\alpha\delta}\delta_{\beta\gamma}-\delta_{\alpha\beta}\delta_{\gamma\delta}\,,
\label{Eq: spin identity}
\end{align}
the boson exchange contributions to $U_k$ in \cref{equ:action_int} lead to the term 
\begin{align}
\lambda_k^{(a)} = 2 \hat{G}^{(a)}_k\fn{P_2-P_3} + \hat{G}^{(a)}_k\fn{P_1-P_2}\,.
\label{eq: decomposed four-fermion vertex}
\end{align}

This procedure corresponds to the inversion of the Hubbard-Stratonovich transformation.
The two terms can be associated with two boson exchange processes.
Either one electron with momentum ${\vect p}_2$ and a hole with momentum $-{\vect p}_3$ can form a boson with momentum ${\vect q}={\vect p}_2-{\vect p}_3$.
Or, two electrons with momenta ${\vect p}_2$ and ${\vect p}_4$ are scattered to two electrons with momenta ${\vect p}_1$ and ${\vect p}_3$, with momentum transfer ${\vect q}={\vect p}_1-{\vect p}_2$ carried by the exchanged boson, and similar for hole-hole scattering. 
We may represent the two processes graphically by
\begin{align}
\vcenter{\hbox{\includegraphics[width=70mm]{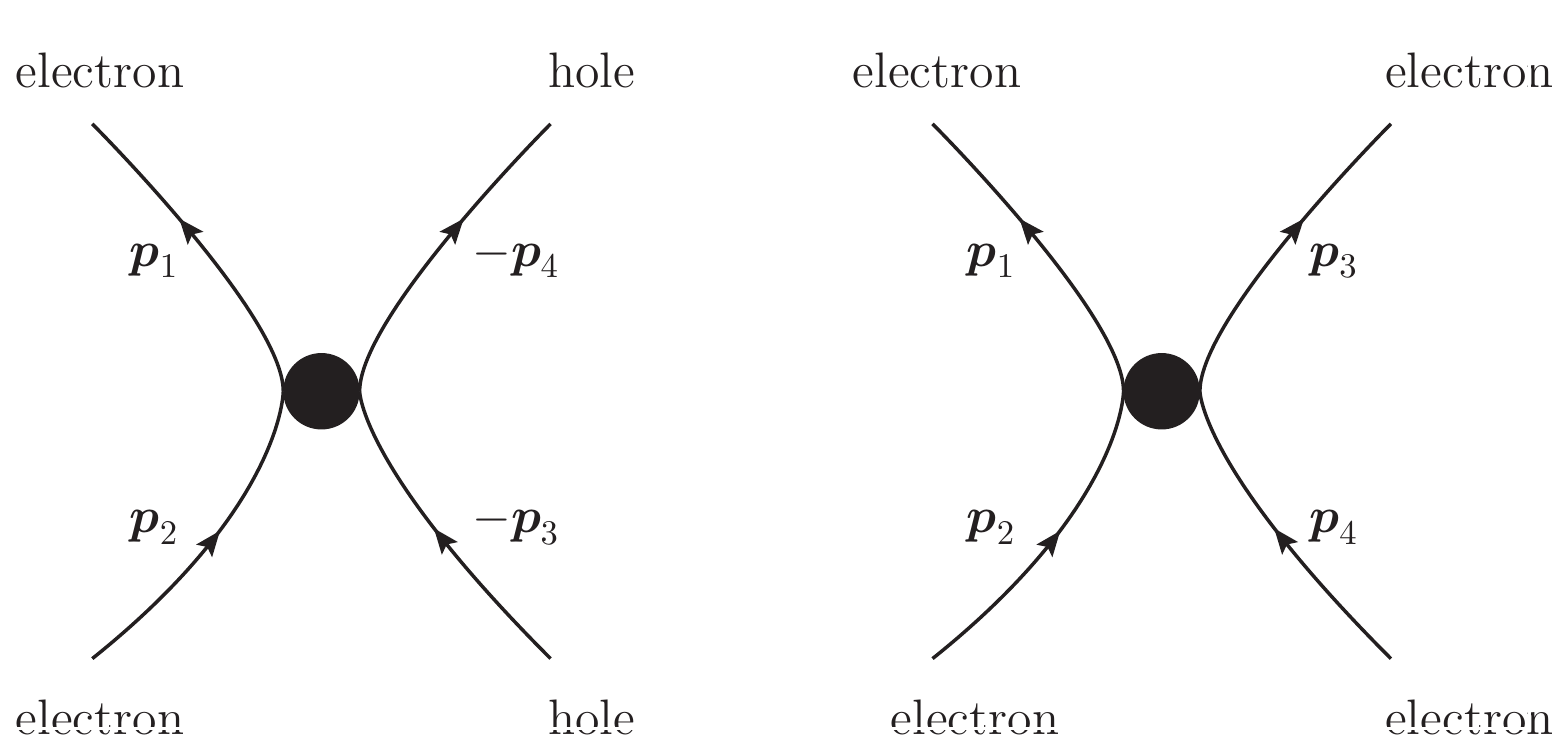}}\vskip0mm}.
\label{eq: diagrammatic fermion scattering process}
\end{align}
According to the momentum conventions in \cref{equ:action_int}, the spin is conserved along each line.
These graphs already account for the reordering of spin indices according to \cref{Eq: spin identity}.
They are different from the Feynman graphs for which the boson-fermion vertex involves $\sigma^j$.

The residual four-fermion interaction after subtraction of the boson exchange contribution is given by
\begin{align}
\lambda_k^{(\psi)} = U_k - \lambda_k^{(a)}\,,
\end{align}
where the superscript $(\psi)$ stands for the residual pure four-fermion interaction corresponding to the last term on the right-hand side of \cref{eq: diagrammatic bosonisation process}.
We are interested in the relative size of $\lambda_k^{(\psi)}$ as compared to $\lambda_k^{(a)}$.
More precisely, we compare the momentum dependent part $\Delta\lambda_k^{(\psi)}$ of $\lambda_k^{(\psi)}$ to $U_k$, subtracting a momentum independent part:
\begin{align}
\Delta\lambda_k^{(\psi)}\fn{P_i} = \lambda_k^{(\psi)}\fn{P_i} - U^{(0)}_k \,.
\end{align} 
The constant part $U^{(0)}_k$ could also be incorporated in $\lambda_k^{(a)}$, where it would amount to an additive constant in the propagator, $\Delta \hat{G}^{(a)}_k = U^{(0)}_k/3$.
It seems natural, however, to keep it as part of the residual four-fermion interaction $\lambda_k^{(\psi)}$ as in Refs.\,\cite{Friederich:2010hr,Friederich:2010zn}.

As an example of a momentum configuration of the four-fermion coupling, we take the momentum configuration $(P_1,P_2,P_3,P_4)=(P,P,-P,-P)$, which corresponds to the ``crossed particle-hole'' channel in accordance with~\cite{Salmhofer:2001tr,Metzner:2011cw}.
For this configuration we obtain
\begin{align} 
\lambda_k^{(a)}(P) &=\frac{2}{m_{k}^2+A_k[(2p_x-\pi)^2+(2p_y-\pi)^2]}\nn[1ex]
&\qquad
+\frac{1}{m_{k}^2+2\pi^2 A_k}\,.
\label{eq: structure of bosonised four-fermion coupling constant}
\end{align}
The upper left part in \cref{fig:ff_coupling_IR_main_figure} shows $U_\text{CPH}\fn{P}=U_k\fn{P,P,-P,-P}$ with $P=(0,p_x,p_y)$ as a function of $(p_x,p_y)$.
We observe pronounced peaks of $U_k\fn{p_x,p_y}$ at $(p_x,p_y)=(\pm \pi/2,\pm \pi/2)$. 
At these values of the momenta one has $[(2p_x-\pi)^2+(2p_y-\pi)^2]=0$, such that $\lambda_k^{(a)}\fn{p_x,p_y}$ has a maximum.
The upper right part in \cref{fig:ff_coupling_IR_main_figure} shows $\lambda_k^{(a)}\fn{p_x,p_y}$.
We find good agreement with $U_k\fn{p_x,p_y}$ -- the four-fermion vertex is indeed well described by a boson exchange.
In order to determine $m_k^2$ and $A_k$, we have performed a three parameter fit of $U_k$ with the ansatz 
\begin{align}\label{eq: bosonised four-fermion coupling constant}
U_k\fn{\bvec p} &= \lambda_k^{(a)}\fn{\bvec p} + U^{(0)}_k \nonumber \\
&= 2 (m_{k}^2+A_k[2\vect{p}-\vect{\pi}]^2)^{-1} + \lambda^{(0)}_k \,.
\end{align}
At $k_\text{IR} \approx 0.9t$ we find
\begin{align}
&m_k^2=0.32\,,&
&A_k=0.53\,,&
&U_k^{(0)}=1.08\,.&
\end{align}
The second part in \cref{eq: structure of bosonised four-fermion coupling constant} only contributes a small part to the constant $\lambda^{(0)}$ obtained from the fit,
\begin{align}
\lambda_k^{(0)}-U_k^{(0)}=\frac{1}{m_k^2+2\pi^2A_k}=0.09\,.
\end{align}
The residual four-fermion interaction $\Delta\lambda_k^{(\psi)}=U_k-\lambda_k^{(b)} - U^{(0)}_k$ is displayed in the lower left panel of \cref{fig:ff_coupling_IR_main_figure}.

Other momentum configuration cases will be investigated in \cref{sect: Partial bosonisation} (see \cref{fig:ff_coupling_IR_PHD,fig:ff_coupling_IR_PP}).
The results support the overall effectiveness of partial bosonisation. 
In the following sections, we introduce our setups for the Hubbard model and the FRG and then explain in a more concrete way how we fit the boson exchange contributions to $U_{k_\text{IR}}(P_1,P_2,P_3,P_4)$.

\section{Hubbard model} \label{sec:hubbard_model}
The Hubbard model describes non-relativistic fermions on a lattice, e.g., as a toy model for electrons in Mott insulators.
For a 2D square lattice, the microscopic action of the single-band fermionic Hubbard model in momentum space is given by
\begin{align} \label{equ:action_bare}
  S(\psi^{\dag},\psi) &= \sum_Q \psi^{\dag}_{\alpha}(Q) \left( \imaginaryi \omega_Q + \xi_{\vect{q}} \right) \psi^{}_{\alpha}(Q) \nonumber \\[1ex]
  &\phantom{=\;} 
  + \frac{U}{2} \sum_{P_i} \psi^{\dag}_{\alpha}(P_1) \psi^{}_{\alpha}(P_2) \psi^{\dag}_{\beta}(P_3) \psi^{}_{\beta}(P_4) \nonumber \\[1ex]
  &\phantom{=\;+\,} \qquad\qquad \times \delta(P_1-P_2+P_3-P_4) \,.
\end{align}
Here, we have introduced the short hand notations $Q \colonequ (\omega_Q,\vect{q})\equiv(\omega_Q,q_x,q_y)$ and
\begin{align}
  \delta(P-Q) &\colonequ \frac{(2\pi)^2}{T} \delta(\omega_P- \omega_Q) \delta^{(2)}(\vect{p} - \vect{q}) \;, \nonumber \\[1ex]
  \sum_Q &\colonequ T\sum_{n \in \mathbb{Z}}\int\limits^\pi_{-\pi}\frac{\mathrm{d}^2 q}{(2\pi)^2}\,.
\end{align}
The fermionic fields
\begin{align}
  \psi(Q) &=  \left( \psi^{}_{\uparrow}(Q), \psi^{}_{\downarrow}(Q) \right)^{T} 
\end{align}
describe fermions (electrons) on a square lattice which follow the momentum-space dispersion relation:
\begin{align}
\label{the dispersion relation for electron}
 \xi_{\vect{q}}
   =- \mu - 2 t& \left( \cos(q_x) + \cos(q_y) \right)\,.
\end{align}
Here, $t$ 
denotes the the nearest-neighbour
hopping parameters, and $\mu$ is the chemical potential associated with the doping level. 
In this work, we do not take into account the next-to-nearest-neighbour hopping.
The fermionic Matsubara frequencies at a finite temperature $T$ are given by $\omega_Q = 2 \pi ( n_Q+ \frac{1}{2} ) T$.
All components of the generalised momentum $Q$ are measured in units of the lattice spacing $a$, i.e., $\vect{q}\to \vect{q}/a$.
In this work, we always set $a=1$.

The action \eqref{equ:action_bare} is invariant under global U(1) and SU(2) rotations.
The U(1) symmetry corresponds to charge (or particle number) conservation, while the SU(2) symmetry reflects the spin-rotation invariance of the system.
In addition, the Hubbard model is invariant under simultaneous rescalings of the imaginary time $\tau$ and the corresponding inverse rescalings of the parameters $U$, $T$, and $t$.
For this reason, we express these quantities in units of the nearest-neighbour hopping $t$ and consequently set $t=1$ most of the time.
Our numerical results employ $\mu=0$, $T=0.1$, and $U/t=3$.

The energy dispersion relation $\epsilon_{\vect{q}}=\mu+\xi_{\vect{q}}$ satisfies the nesting
property, i.e.,
\al{
\label{eq: nesting property}
\epsilon_{\vect{q}+\vect{\pi}}=-\epsilon_{\vect{q}}\,, } where
$\vect{\pi}=(\pi,\pi)$ is the nesting vector.  In particular, for
$\mu=0$ the half-filled band is realised and the fermion surface
becomes a perfect square, the so-called umklapp surface.
In this case the particle-hole channels are strongly enhanced.  More
specifically, for the momentum configurations
$\vect{p}_1-\vect{p}_2=\pm\vect{\pi}$ and
$\vect{p}_1-\vect{p}_4=\pm\vect{\pi}$, quantum fluctuations
corresponding to bubbles of electron-hole pairs become strong, so
that antiferromagnetic ordering (electron-hole condensate)
takes place for any $U>0$. 
We discuss this in more detail in
\cref{subsec: Flow of the four-fermi coupling and antiferromagnetic
  ordering}.

\section{Functional renormalisation group} \label{sec:frg}
In this section, we introduce the functional renormalisation group (FRG).
We then make an ansatz for the effective average action in order to solve the FRG equation.
In this work, we take into account only the flow of the four-fermion interaction.
We also describe our choice of the fermion regulator.

\subsection{Flow equation}
The FRG is a formulation of Wilson's RG in the context of quantum field
theory.
According to Wilson, renormalisation should be thought of as a sequence of coarse-graining steps, each corresponding to integrating out quantum fluctuations in an infinitesimal momentum shell $k-\delta k <p<k$, where $k$ denotes an IR cutoff scale.
This process can be formulated as a functional differential
equation for a scale-dependent effective action without any
approximation.
In this work, we employ the exact flow equation for
the scale-dependent one-particle-irreducible (1PI) effective action
(or effective average action) $\Gamma_k$
\cite{Wetterich:1992yh}, which reads as
\begin{align} \label{equ:wetterich_equation}
\partial_k \Gamma_k = \frac{1}{2}\Tr \,\left[ \frac{1}{\Gamma^{(2)}_k+R_k }\, \partial_k R_k \right].
\end{align}
Here, $\Tr$ denotes a generalised functional trace over field space and all internal spaces such as frequencies, momenta, and spins.
$\Gamma^{(2)}_k$ stands for the matrix of second functional derivatives of $\Gamma_k$ with respect to the fields, i.e., the inverse propagator.
The regulator $R_k$ implements the coarse-graining within momentum integrations of \cref{equ:wetterich_equation} such that low-momentum modes $p<k$ are suppressed, and thus only high-momentum modes $p>k$ are effectively integrated out.
In other words, the regulator in particular has to satisfy the conditions
\begin{align}
\label{eq: conditions for regulator}
&\lim_{k\to0}R_k(p)=0\,,&
&\lim_{k\to \Lambda \to \infty}R_k(p)=\infty\,.&
\end{align}
The first condition implies that for $k\to0$ one obtains the fully dressed effective action $\Gamma_{k=0}=\Gamma$.
From the fact that the regulator behaves as a mass term for a finite IR scale $k$, the second condition is needed to suppress all quantum fluctuations in the high energy limit, so that the initial condition for the flow is given by fixing the microscopic action $S$ at the UV scale $k = \Lambda$, i.e., $\Gamma_\Lambda = S$.
For more details on the FRG and its applications, see e.g. Refs.~\cite{Berges:2000ew,Pawlowski:2005xe,Gies:2006wv,Delamotte:2007pf,Kopietz:2010zz,Braun:2011pp,Metzner:2011cw}.

\subsection{Effective average action}
Although the flow equation~\eqref{equ:wetterich_equation} is exact, it generally cannot be solved exactly.
In general, the effective action $\Gamma_k$ obeying the flow equation \eqref{equ:wetterich_equation} involves an infinite number of effective operators compatible with the symmetries of the theory.
Therefore, we need to truncate our theory space in order to be able to do computations in practice.
To this end, we make an ansatz for the effective action $\Gamma_k$ and choose suitable projections to describe its flow in terms of flowing couplings.
Here, we make the simple ansatz
\begin{align} \label{equ:action_eff}
  \Gamma_k&= \sum_Q \psi^\dagger_\alpha(Q) \tilde P^{\text{f}}_k(Q) \psi^{}_\alpha(Q)
+ \frac{1}{2}\sum_{P_i} U_k(P_1,P_2,P_3,P_4)\nn
&\phantom{= \sum_Q \psi^\dagger_\alpha(Q) }
\times \psi^\dag_\alpha(P_1)\psi^{}_\alpha(P_2) \, \psi^{\dag}_\beta(P_3) \psi^{}_\beta(P_4) \nn
&\phantom{= \sum_Q \psi^\dagger_\alpha(Q)\tilde P^{\text{f}}_k(Q)}
\times\delta(P_1-P_2+P_3-P_4) \;,
\end{align}
where $\tilde P^{\text{f}}_k(Q) = \imaginaryi\, \omega_Q + \xi_{\vect{q}}$ is the inverse fermion propagator, with $\omega_Q$ the Matsubara frequency and $\xi_{\vect{q}}$ the dispersion given in \cref{the dispersion relation for electron}.
In addition to the cutoff scale, the four-fermion coupling constant in the effective action \eqref{equ:action_eff} depends on external momenta.

\subsection{Fermion regulator}
\label{sec:fermion_regulator}
There are various possible choices for the regulator $R^{\text{f}}_k(Q)$ which obey the conditions~\eqref{eq: conditions for regulator}.
Here, we focus on the fact that the finite-temperature effect avoids the existence of zero modes in the fermionic propagator, i.e., $\omega_Q=\pi T$ for $n=0$.
Hence, the temperature itself plays the role of a regulator.
For fermions we thus use an effective temperature cutoff~\cite{Baier:2003ex,Baier:2003fw}:
\begin{align} \label{equ:regulator_fermion}
  R^{\text{f}}_k(Q) &= \imaginaryi \omega_Q \left( \frac{T_k}{T} - 1 \right)= 2 \pi \imaginaryi \left(n+\frac{1}{2}\right) (T_k-T) \;.
\end{align}
With this cutoff, the temperature $T$ in the Matsubara frequency is replaced by the scale-dependent effective temperature $T_k$.
Here a general class of the scale-dependent effective temperature $T_k$ is given by
\al{
\left(T_k\right)^p =T^p+\left( \frac{k^2}{\pi}\right)^p\,,
}
with $p$ an integer larger than 2.
In the long wavelength limit $k^2\ll T$, the fermionic effects should be suppressed in order for the bosonic picture to be relevant.
For this reason, $p$ has to be a sufficiently large value.
In this work, we choose $p=4$,
for which the scale-dependent effective temperature is given as 
\begin{align} \label{equ:temp_eff}
  T_k &= \left( T^4 + \left( \frac{k^2}{ \pi } \right)^4 \right)^\frac{1}{4} \;.
\end{align}
For $k^2\gg T$, one has $T_k\approx k^2/\pi$, such that the regulated inverse propagator behaves as $P_k\approx \imaginaryi (2n+1)k^2+\xi_{\vect{q}}$, so that the zero mode is regulated by $k^2$, while for $k^2\ll T$ the finite temperature $\pi T$ regulates the zero mode.
This is compatible with the fact that in non-relativistic systems the lowest order of the kinetic energy is proportional to $\vect{p}^2$, namely, one can interpret the behaviour $T_k\sim k^2$ for $k^2\gg T$ as a regularisation of the kinetic term of the fermionic field.
Our numerical results are near $T=0$ where $T_k\approx k^2/\pi$.

\section{Flow of the four-fermion coupling}
\label{Sec: solving pure fermionic equation}
In this section we present the flow equation of the four-fermion vertex and our approach to solving it.
In addition, we investigate the resulting momentum dependence of the four-fermion coupling in the IR in the purely fermionic setting.

Using the flow equation \eqref{equ:wetterich_equation} and projecting onto the momentum-dependent four-fermion coupling $U_k(P_1,P_2,P_3,P_4)$ in the effective action \eqref{equ:action_eff}, we obtain the beta function
\begin{align}
\label{eq: flow equation for four-fermion vertex}
\p_k U_k(P_1,P_2,P_3,P_4)={\mathcal T}_\text{PP}+{\mathcal T}_\text{DPH}+{\mathcal T}_\text{CPH}\,,
\end{align}
where ``PP," ``DPH,'' and ``CPH'' denote the particle-particle, direct particle-hole, and crossed particle-hole channel, respectively.
We present the details of our projection scheme and the explicit forms of the flow diagrams $\mathcal T_i$ in \cref{sec:flow_equations}.
At the initial scale $k=\Lambda$, we set the four-fermion coupling to a constant, i.e. $U_\Lambda(P_1,P_2,P_3,P_4) = U$.
The flow equations hold for arbitrary temperature $T$. Our numerical results are evaluated at $T=0.1$.

\subsection{\texorpdfstring{$N$}{N}-patching and frequency dependence}
The  four-fermion vertex $U_k$ is a function of four generalised momenta $P_i = ( \omega_{i}, \vect{p}_i)$, each consisting of a Matsubara frequency $\omega_i$ and a two-dimensional momentum $\vect{p}_i$, giving a dependence on 12 parameters in total.
Conservation of energy and momentum allows us to eliminate one generalised momentum, e.g., $P_4$, which, however, still leaves us with nine parameters in total.
To further reduce the computational effort of our study, we thus make use of two main simplifications, which we explain in the following.

In the low-energy limit the most relevant effects would come from the dynamics of electrons on the Fermi surface.
In other words, it is expected that the scaling dimensions of the four-fermion couplings with external energies and momenta perpendicular to the Fermi surface are negative at the fixed point associated to the antiferromagnetic phase transition, i.e., these couplings are irrelevant parameters in the sense of Wilson's RG, while the couplings corresponding to tangential external momenta are relevant parameters.
Hence, as a first approximation, we neglect the external frequency dependence of the four-fermion coupling.
This makes it possible to analytically perform the Matsubara sums occurring in the flow diagrams on the right-hand side of equation~\eqref{eq: flow equation for four-fermion vertex}.
We show the explicit computations in \cref{sec:matsubara_sums}.

In addition, we approximate the momentum dependence of the four-fermion vertices occurring on the right-hand side of equation~\eqref{eq: flow equation for four-fermion vertex} by projecting the
two dimensional momenta onto the Fermi surface, i.e.,
\al{
U_k\fn{P_1,P_2,P_3,P_4}\approx U_k\fn{{\bvec p}_{1F},{\bvec p}_{2F},{\bvec p}_{3F}}\,,
}
where $\vect{p}_{i F}$ denotes the projections of $\vect{p}_{i}$ onto the Fermi surface, and where we have explicitly used momentum conservation to eliminate $\vect{p}_4$.
This allows us to still resolve the full momentum dependence of the four-fermion coupling in particular momentum channels, while significantly reducing computational effort by only feeding back a coupling depending on three parameters.

Furthermore, to numerically solve the flow of $U_k$ in practice, we cannot evaluate $U_k$ for continuous parameters, but instead employ a multi-dimensional grid of discrete parameters.
In particular, we employ an $N$-patching scheme~\cite{1997ZPhyB.103..339Z} for the four-fermion vertices fed back on the right-hand side of the flow equation of $U_k$.
This is achieved by dividing the Brillouin zone into $N$ patches for which momenta are parametrised by indices corresponding to the position of the patch, e.g., $\vect{p}_{1F}= \vect{p}_{n_1}$.
All momenta in the same patch labeled by $n_i$ are projected onto the same momentum on the Fermi surface, at the center of the patch.
This way, in order to feed back the four-fermion vertex, we only need to keep track of a three-dimensional grid of couplings
\begin{align}
  U_k(\vect{p}_{1F}, \vect{p}_{2F}, \vect{p}_{3F} ) &\longrightarrow U_k( n_1, n_2, n_3 )\,.
\end{align}
The results presented in the following use $N=24$.

\subsection{Antiferromagnetic ordering}
\label{subsec: Flow of the four-fermi coupling and antiferromagnetic ordering}
As mentioned in \cref{sec:hubbard_model}, at half filling with $\mu = 0$, antiferromagnetic ordering forms at sufficiently low temperature.
A tendency to antiferromagnetism can be seen from the beta function for $U_k$.
For the momentum configurations ${\vect  p}_1-{\vect p}_2=\vect{\pi}$ (direct particle-hole channel) and ${\vect p}_1-{\vect p}_4=\vect{\pi}$ (crossed particle-hole channel), the kernel of threshold functions defined by $L_k(Q,Q-P)=\p_k (G^\text{f}_k(Q) G^\text{f}_k(Q-P))$ does not depend on external momenta thanks to the nesting property \eqref{eq: nesting property}, i.e. $L(Q,Q-P)=L(Q,Q)$.
Within the $N$-patch approximation the dispersion \eqref{the dispersion relation for electron} can be expanded around the Fermi surface,
\al{
\xi_{{\vect q}}\approx \xi_{{\vect q}_F}+{\vect v}_F|_{{\vect q}={\vect q}_F}\cdot ({\vect q}-{\vect q}_F)\,,
}
where ${\vect v}_F=\nabla_{\vect q}\xi_{\vect q}$ is the Fermi velocity.
At half filling one has $\epsilon_{{\vect q}_F} = \xi_{\vect{q}_F} + \mu =0$.
The saddle points $(\pm\pi,0)$ and $(0,\pm\pi)$ have a vanishing Fermi velocity ${\vect v}_F={\vect 0}$ which causes the so-called van Hove singularities in the density of states.
Another solution to ${\vect v}_F={\vect 0}$ is the nesting vector ${\vect \pi}$.
Therefore, quantum fluctuations of particle-hole bubbles, $L\fn{Q,Q}$, for the nesting external momenta enhance around the vanishing Fermi velocity ${\vect v}_F={\vect 0}$ because of the umklapp scattering with momentum transfer $\bvec \pi$ and the nesting property.
This strong enhancement is expected to generate antiferromagnetic order at sufficiently low $T$.

We present the results for $U_k$ evaluated on the crossed and direct particle-hole channels, $U_{\text{CPH}}$ and $U_{\text{DPH}}$, in \cref{fig:ff_coupling_IR_main_figure,fig:ff_coupling_IR_PHD}. 
As can be seen in the top left plots of the figures, the particle-hole channels exhibit strongly enhanced peaks at momenta $\vect{p} = (\pm\pi/2,\pm\pi/2)$, which drive antiferromagnetic ordering effects. 
This behaviour of the four-fermion coupling constant is driven by the umklapp and nesting processes mentioned above.
We can interpret this as a signal for the breaking of SU(2) spin rotation invariance in the system and as an emergence of antiferromagnetic ordering.

\section{Partial bosonisation}
\label{sect: Partial bosonisation}
As explained in the last section, we integrate down the flow of the four-fermion coupling until we reach a scale close to the onset of antiferromagnetic ordering.
There the divergence of nesting peaks in the particle-hole channels signals the transition to an antiferromagnetic phase, and we can no longer follow the flow of the four-fermion coupling in the purely fermionic setting.
As discussed in \cref{subsec: Flow of the four-fermi coupling and antiferromagnetic ordering}, the nesting peaks are indeed the most prominent features in the particle-hole channels.
In order to follow the flow of the four-fermion coupling into the symmetry-broken regime, we need to bosonise its corresponding channels.

In this section, we first present our results, which show that the non-trivial shape of the four-fermion coupling in different momentum channels can be well described by boson exchange processes, parametrised by a few fitting parameters. 
After a short review of partial bosonisation via the Hubbard-Stratonovich transformation, we explain our ansatz for the effective action to capture particle-hole or particle-particle scatterings in a bosonic language.
We then attempt to bosonise the four-fermion coupling into an antiferromagnetic order boson and an additional boson which can be associated to $s$-wave Cooper pairing. We also investigate the shape and relative size of the residual four-fermion coupling.

\subsection{Different momentum channels} \label{sec:momentum_channels}
In order to resolve the momentum dependence of the four-fermion coupling in more detail, we concentrate on three specific momentum channels, for which we employ two-dimensional grids of discrete momenta $P = ( 0, \vect{p})$ with $\vect{p} = ( p_x, p_y )$.
In particular, we investigate the momentum dependence of the four-fermion coupling for the following momentum configurations:
\begin{align}
  U_{\text{CPH}}(P) &\colonequ U_k(P,P,-P,-P)\,, \nonumber\\[0.5em]
  U_{\text{DPH}}(P) &\colonequ U_k(P,-P,-P,P)  \,, \nonumber \\[0.5em]
  U_{\text{PP}}(P) &\colonequ U_k(P,P,P,P)\,.
  \label{Eq: four-fermion vertex for different channels}
\end{align}
These momentum configurations have in common that for each of them specific channel momenta take a maximal value $2 P$, namely,
 $P_\text{CPH}\colonequ P_1 - P_4$, $P_\text{DPH}\colonequ P_1 - P_2$, and $P_\text{PP}\colonequ P_1 + P_3$, while many other sums or differences of momenta are zero.
They can thus be associated to the crossed and direct particle-hole channels, and the particle-particle channel.

\begin{figure*}
  \includegraphics[width = \linewidth]{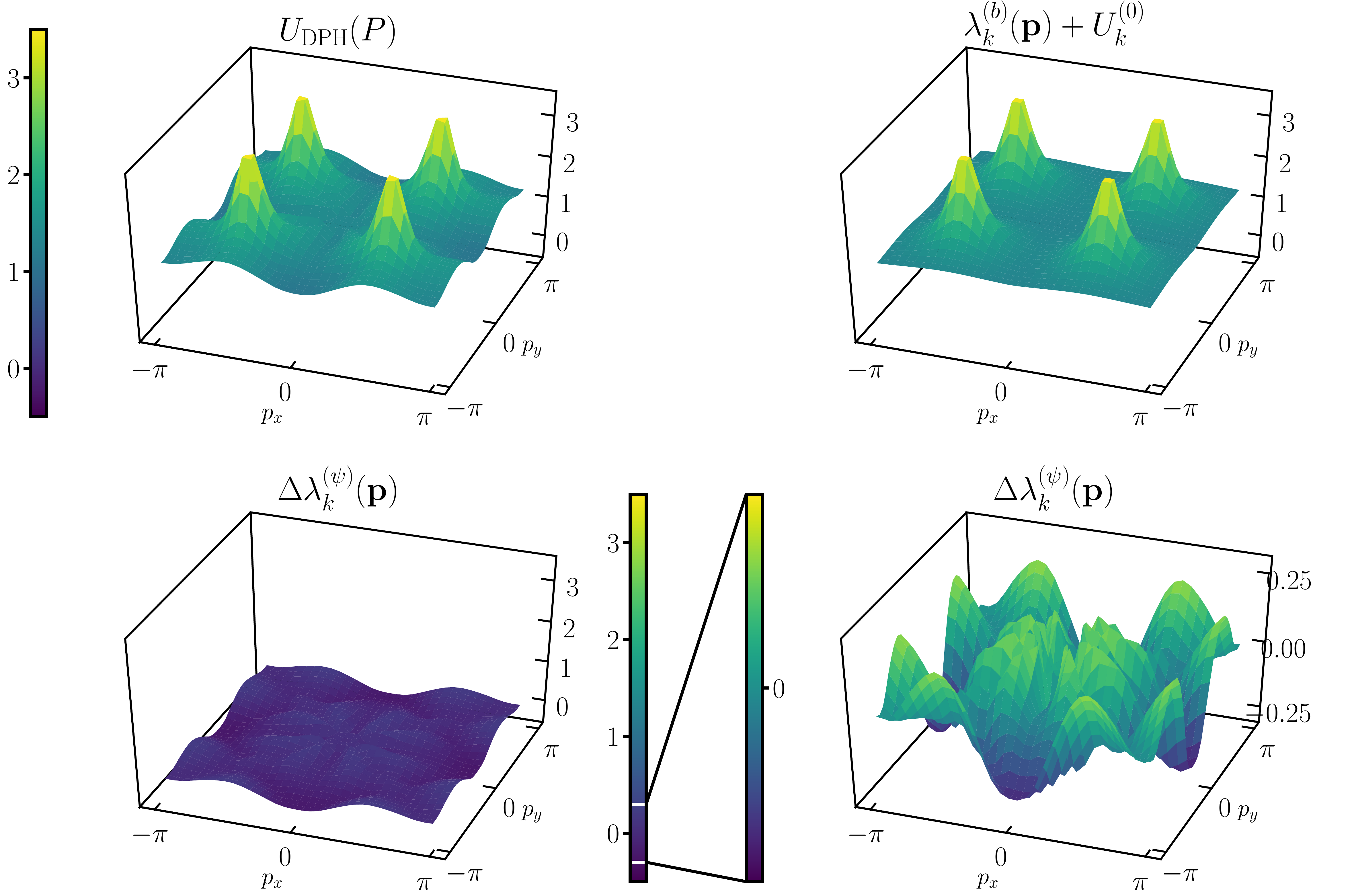}
  \caption[Four-fermi coupling in the IR]{
  Four-fermion coupling $U_k(P,-P,-P,P)=U_\text{DPH}\fn{P}$, $P=(0,p_x,p_y)$, where at one-loop level only the direct particle-hole channel (DPH) contributes.
  Parameters are the same as for \cref{fig:ff_coupling_IR_main_figure}.
  The numerical result shown in the top left panel is dominated by nesting peaks located at $(\pm\pi/2,\pm\pi/2)$.
  The top right panel shows a bosonic exchange propagator of the form $\lambda_{\text{DPH}}^{(b)}(\vect{p}) + U_k^{(0)} = ( m_k^2+A_k [2 \vect{p} - \vect{\pi}]^2)^{-1} + \lambda^{(0)}_\text{DPH}$ fitted to the result.
  We find that $U_\text{DPH}\fn{P} \approx \lambda_\text{DPH}^{(b)}\fn{\vect p} +U_k^{(0)}$ to high accuracy.
  The bottom panels show the the momentum dependence of the residual four-fermion vertex as difference between $U_\text{DPH}\fn{P}$ and $\lambda_\text{DPH}^{(b)}\fn{\vect p} + U_k^{(0)}$, where the bottom right panel is a zoomed in version of the one on the bottom left.
Differences between $U_\text{DPH}\fn{P}$ and $\lambda_\text{DPH}^{(b)}\fn{\vect p}+ U_k^{(0)}$ are overall about one order of magnitude smaller than the peaks.
  }
  \label{fig:ff_coupling_IR_PHD}
\end{figure*}

Boson exchanges affect the four-fermion vertex in different momentum channels differently.
For example, the exchange of an antiferromagnetic boson contributes only a small constant in the particle-particle channel.
With \cref{eq: decomposed four-fermion vertex} the antiferromagnetic boson exchange vertex is given by
\begin{align}
\lambda_k^{(a)}=\frac{n_1}{m_k^2+A_k[2{\bvec p}-\vect{\pi}]^2}+\frac{n_2}{m_k^2+2\pi^2 A_k}\,,
\label{eq: fitting ansatz for four-fermion coupling no2}
\end{align}
with $n_1$ and $n_2$ integers, $n_1+n_2=3$, given in \cref{Table: fitting results}.
The result of the fit for $U_\text{DPH}$ is shown in \cref{fig:ff_coupling_IR_PHD}, similar to the one for $U_\text{CPH}$ in \cref{fig:ff_coupling_IR_main_figure}.
The fitted parameters $m_k^2$, $A_k$, and $\lambda_k^{(0)}$ are shown in \cref{Table: fitting results}.

\begin{table}
\begin{center}
\begin{tabular}{ c c c c c c }
\toprule
    \makebox[0.5cm]{}  &  \makebox[2.5cm]{$(P_1,\,P_2,\,P_3,\,P_4)$} & \makebox[1.8cm]{$\lambda_k^{(b)}\fn{n_1,n_2}$} & \makebox[0.9cm]{$m_k^2$} & \makebox[0.9cm]{$A_k$}  & \makebox[0.9cm]{$U_k^{(0)}$}  \\
\midrule
CPH & $(P,\,P,\,-P,\,-P)$ & $(2,\,1)$ & $0.32$ & $0.53$ & $1.08$  \\
DPH & $(P,\,-P,\,-P,\,P)$ & $(1,\,2)$ & $0.41$ & $0.69$ &$1.08$  \\
PP & $(P,\,P,\,P,\,P)$ & $(0,\,3)$ & --- & --- & --- \\
\bottomrule
\end{tabular}
\caption{
\label{Table: fitting results}
Numerical results for fitting the ansatz \eqref{eq: fitting ansatz for four-fermion coupling no2} in different momentum configurations. An exchange of an antiferromagnetic boson $a$ leads to a momentum-dependent contribution to the particle-hole channels, multiplied by the integer $n_1$, as well as a constant contribution in all three channels, multiplied by $n_2$. We can extract the fit parameters $m_k^2$ and $A_k$ parametrising the boson propagator from the momentum-dependent part, while the constant part together with an additional constant part of the four-fermion interaction $U_k^{(0)}$ leads to a total constant shift $\lambda_k^{(0)}$. Since an antiferromagnetic boson exchange only contributes an overall constant to the particle-particle channel, we cannot perform the same fit for this channel. }
\end{center}
\end{table}

In an ideal situation of a pure boson exchange interaction, the same parameters $m_k^2$ and $A_k$ should describe $U_k$ for various momentum configurations.
We observe that the values of $m_k^2$ and $A_k$ extracted from the fits in the crossed and direct particle-hole channels are indeed of similar size, confirming the possibility of a partially bosonised description.
In particular, \cref{eq: fitting ansatz for four-fermion coupling no2} implies that the amplitude of the antiferromagnetic peaks in the CPH channel is twice the amplitude of the peaks in the DPH channel.
This is approximately realised by the computed four-fermion interactions.
For the peak locations $P_0$ we find $U_{\text{CPH}}(P_0) \approx 6.87$, and $U_{\text{DPH}}(P_0) \approx 3.46$.
The absence of an antiferromagnetic enhancement in the PP channel is also realised by the computed four-fermion interaction, as we will see in more detail in \cref{sec:propagator_fit_results}.

\subsection{Bosonisation of the four-fermion interaction}
\label{sec:four-fermi_bosonisation}

\begin{figure*}
  \includegraphics[width = \linewidth]{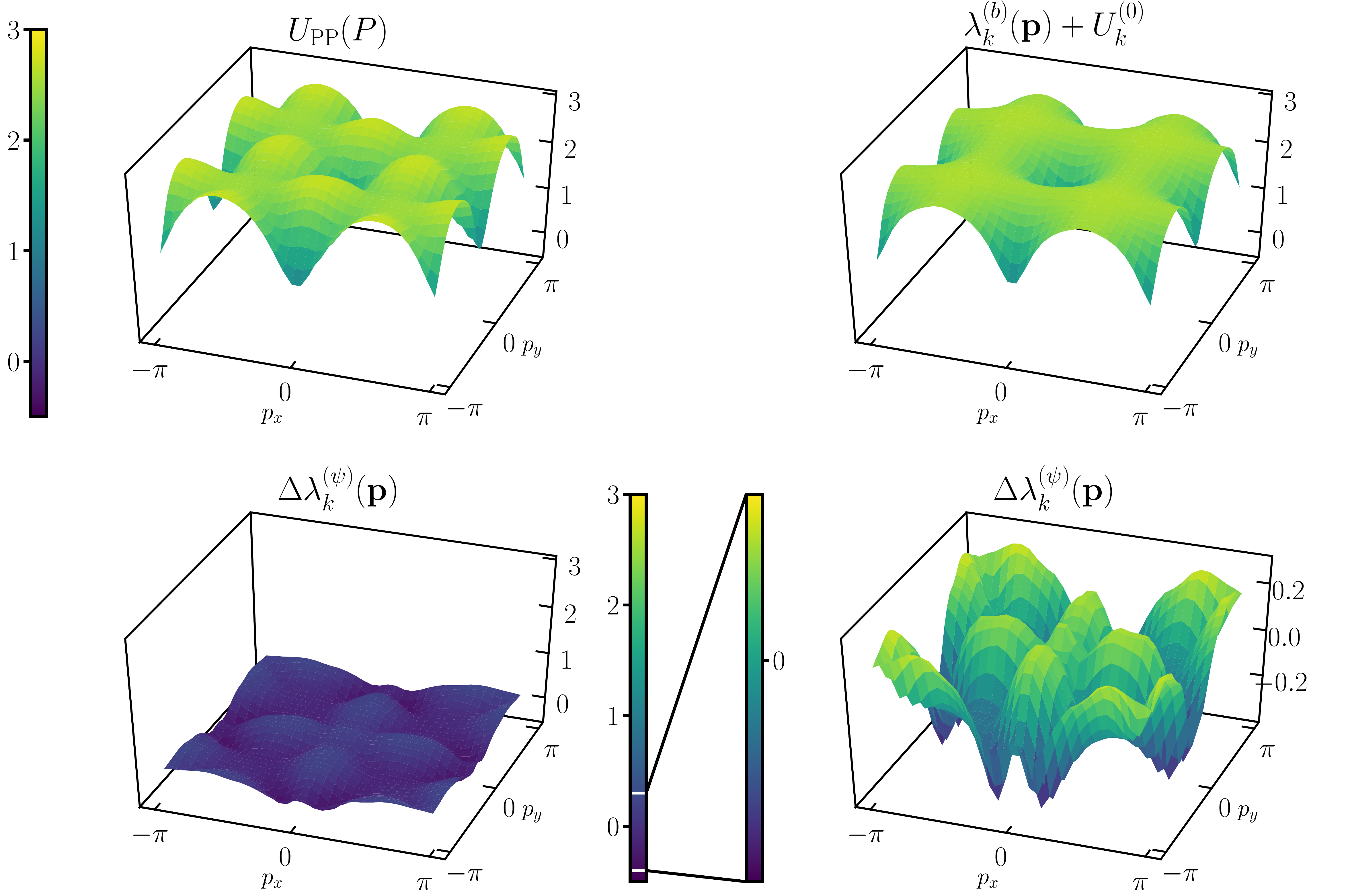}
  \caption[Four-fermi coupling in the IR]{
  Four-fermion coupling $U_k(P,P,P,P)=U_\text{PP}\fn{P}$, $P=(0,p_x,p_y)$, where at one-loop level only the particle-particle channel (PP) contributes.
The numerical result shown in the top left panel is qualitatively different from the one in the particle-hole channels.
  Instead of nesting peaks, we observe a suppression of $U_k$ at $p_x=0,\,\pm \pi$ and $p_y=0,\,\pm \pi$.
  The top right panel shows a bosonic exchange propagator of the form $\lambda_{\text{PP}}^{(b)}(\vect{p}) + U_k^{(0)} = ( m_k^2+A_k [2 \vect{p}]^2)^{-1} + \lambda^{(0)}_\text{PP}$ fitted to the result. 
  We find that $U_\text{PP}\fn{P} \approx \lambda_\text{PP}^{(b)}\fn{\vect p}+U_k^{(0)}$ is a good approximation.
  The bottom panels show the difference between $U_\text{PP}\fn{P}$ and $\lambda_\text{PP}^{(b)}\fn{\vect p}+U_k^{(0)}$, where the bottom right panel is a zoomed in version of the one on the bottom left.
  We find that differences between $U_\text{PP}\fn{P}$ and $\lambda_\text{PP}^{(b)}\fn{\vect p}+U_k^{(0)}$ are much smaller than the peaks.
  }
  \label{fig:ff_coupling_IR_PP}
\end{figure*}

The exchange of an antiferromagnetic boson is not the only possibility. Other collective degrees of freedom, such as electron pairs~\cite{1993Sci...261..337C}, ferromagnetic bosons~\cite{1991JPCM....3.4917V}, or density waves~\cite{SCALAPINO1995329} may play an important role in certain regions of the phase diagram of the Hubbard model.
See also refs.\,\cite{Baier:2003ex,Friederich:2010zn,Friederich:2010hr} in the context of partial bosonisation.
Each would contribute in specific momentum channels.
Partial bosonisation has to account for all the various possibilities, as well as for interactions where two (or even more) collective modes play an important role.

Let us start with a short discussion of the basic notion of bosonisation in a fermionic system.
The Hubbard-Stratonovich transformation allows us to rewrite the purely fermionic action \eqref{equ:action_bare} as a Yukawa system such that
\begin{align} \label{equ:bosonised action_bare}
  &S(\psi^{\dag},\psi,\phi) \nn
  &=\ \sum_Q \psi^{\dag}(Q) \left( \imaginaryi \omega_Q + \xi_{\vect{q}} \right) \psi(Q)
  + \frac{m^2}{2} \sum_{Q}\phi\fn{-Q}\phi\fn{Q}\nn
  &\quad -\sum_{P,Q} h [\psi^{\dag}\fn{P_1} \psi\fn{P_2}\phi\fn{Q}] \delta\fn{Q-P_1+P_2}\,.
\end{align}
At the level of the bare action, the (as of yet unspecified) bosonic field $\phi$ is an auxiliary field, i.e., a nondynamical field.
The action~\eqref{equ:bosonised action_bare} is equivalent to the pure fermionic action~\eqref{equ:action_bare}.
Indeed, plugging the equation of motion for $\phi$ into the action~\eqref{equ:bosonised action_bare}, we get back the original action~\eqref{equ:action_bare}.
The scalar boson mass parameter $m^2$ is related to the four-fermion coupling constant as
\begin{align}
\label{eq: bosonisation at classical level}
U=-\frac{h^2}{m^2}\,.
\end{align}
Here, the inverse scalar mass parameter is regarded as the momentum-independent propagator of the bosonic field.
We can see explicitly that the Yukawa coupling can be eliminated by redefining the bosonic field $\phi\to \phi/h$ in the action \eqref{equ:bosonised action_bare}, i.e., the Yukawa coupling $h$ is a redundant coupling in the bare action. 
Hence, relation~\eqref{eq: bosonisation at classical level} implies that at the classical level the four-fermion vertex is rewritten in terms of an exchange process of the bosonic field, as diagrammatically represented in \cref{eq: diagrammatic bosonisation process}, although as of yet without external momentum dependencies or a residual four-fermion vertex.

The bosonic form~\eqref{equ:bosonised action_bare} corresponds to the exchange of a density wave boson and leads to negative $U$.
We can also use bosonisation with an antiferromagnetic boson in the limit $A_k =0$.
In this case we obtain from \cref{eq: structure of bosonised four-fermion coupling constant}:
\begin{align} \label{equ:bosonisation_classical_af}
  U &= \frac{ 3 h^2 }{ m^2} \,.
\end{align}
The sign is opposite to the exchange of the density wave boson \eqref{eq: bosonisation at classical level}.
There are further possibilities to perform bosonisation along these lines, introducing a substantial amount of arbitrariness in the bosonisation of a point-like four-fermion interaction~\cite{Baier:2003ex}.

Fluctuation effects change the situation profoundly.
They introduce a momentum dependence of $U$.
Now, only particular choices of collective fields can generate by their exchange the dominant features of $U$, provided appropriate momentum-dependent bosonic propagators, and possibly momentum-dependent Yukawa interactions, are chosen.

For a half-filled lattice we expect the occurrence of antiferromagnetic ordering.
The order parameter for this is $\langle \psi^{\dag}\sigma^i \psi\rangle$, which characterises the breaking of SU(2) spin-rotation symmetry.
The collective modes corresponding to such an ordering are in the adjoint representation of SU(2), and we denote them by $m^i\fn{Q}$ here.
A typical feature associated to antiferromagnetic ordering is the occurrence of nesting peaks, for which the sum of incoming external momenta equals $\pm{\bvec \pi}$.
It is therefore useful for the description of antiferromagnetic ordering to introduce collective bosons with a constant shift in the momentum,
\al{ \label{equ:antiferromagnetic_boson_mom_shift}
a^i\fn{Q}=m^i\fn{Q+\Pi}\,,
}
where $\Pi=(0,{\bvec \pi})$.
An expectation value of the antiferromagnetic boson $a^i$ at zero momentum, $\langle a^i(P=0) \rangle \neq 0,$ then corresponds to a homogeneous antiferromagnetic spin density.

\subsection{Pairing bosons}
In addition to the antiferromagnetic bosons $a^i$, we could introduce another collective mode corresponding to Cooper pairs of electrons, i.e., $\varphi\sim \langle \psi^T \varepsilon \psi\rangle$, with $\varepsilon=i\sigma^2$ the totally antisymmetric tensor.
A non-vanishing expectation value of such a collective mode indicates the breaking of U(1) symmetry, which corresponds to charge conservation. 
This type of order is identified with a superconducting phase.
To realise such a phase, however, the electrons should exhibit an effectively attractive force, for which a departure from half-filling may be crucial.
Although Cooper pairing of electrons is not expected to occur in the system at half-filling, we may introduce a collective mode for Cooper pairing.
Partial bosonisation in two channels allows the exploration of more details of the four-fermion interaction and of other regions of the phase diagram.

So far we have encoded the particular momentum dependence of the boson-exchange-induced interaction in the ansatz for the boson propagator $\hat{G}$, [cf. \cref{equ:propagator_boson_af}].
It is often more convenient, however, to account for this momentum dependence by an appropriate momentum-dependent form factor in the fermion-boson interaction.
The inverse boson propagator can then be chosen to have its minimum at zero momentum, as for the antiferromagnetic boson $a^i$ in \cref{equ:antiferromagnetic_boson_mom_shift}.
For the antiferromagnetic boson, the form factor consists of a shift of the boson momentum by $\Pi$.
To avoid confusion, we will in the following refer to the propagators with the minima of their inverses fixed at $\vect{p} = 0$ as $G$ instead of $\hat{G}$.

As a simple truncation for the effective action describing the bosonisation of the four-fermion coupling in the antiferromagnetic or superconducting channels, we use 
\begin{align} \label{equ:bosonised action_effective}
 \Gamma_k 
  =& \sum_Q \psi^{\dag}(Q) \left( \imaginaryi \omega_Q + \xi_{\vect{q}} \right) \psi(Q)
  +\Gamma_k^\text{boson}\nn
  &-\sum_{P_i,Q} h_{a,k}\fn{P_1,P_2,Q}\nn
  & \times [\psi^{\dag}\fn{P_1}\sigma^i\psi\fn{P_2}] a^i\fn{Q}\delta\fn{Q-P_1+P_2+\Pi} \nonumber \\[1ex]
  & - \sum_{P_i,Q} \mathop{\Delta}\left(\frac{ \vect{p}_1 - \vect{p}_2 }{2} \right) h_{\varphi,k}(P_1,P_2,Q) \nonumber \\[1ex]
  & \times \left( [\psi^{T}(P_1)\varepsilon\psi^{}(P_2)] \varphi^{*}(Q)
- [\psi^{\dag}(P_1)\varepsilon\psi^{*}(P_2)] \varphi(Q) \right) \nn[0.5ex]
& \times \delta(P_1 + P_2 - Q)\;.
\end{align}
Here, $a^i$ and $\varphi$ are collective modes describing an antiferromagnetic spin wave and a general Cooper pair, respectively.
We denote by $h_{a,k}$ and $h_{\varphi,k}$ the respective scale- and momentum-dependent Yukawa couplings.
The square brackets denote spin-contracted fermion bilinears. 
The form factor $\mathop{\Delta}(\vect{p})$ specifies the momentum configuration of the Cooper pairing.
Depending on the state of the scattered electrons, the form factor can take the following forms:
\begin{align}
\mathop{\Delta}(\vect{p}) =
\begin{cases}
$1$ & (\text{$s$ wave}) \\[2ex]
\frac{1}{2}\left( \cos\fn{p_x} +\cos\fn{p_y} \right) & (\text{extended $s$ wave})\\[2ex]
\frac{1}{2}\left( \cos\fn{p_x} -\cos\fn{p_y} \right) & (\text{$d_{x^2-y^2}$ wave})\\[2ex]
\sin\fn{p_x}\sin\fn{p_y} & (\text{$d_{xy}$ wave})\,.
\end{cases}
\label{eq: list of form factors}
\end{align}
For other possible form factors, see e.g.~\cite{SCALAPINO1995329}.
In particular, the $d_{x^2-y^2}$-wave configuration may be interesting for the understanding of an occurrence of the superconducting phase in strongly correlated systems since its form factor could describe an effectively attractive interaction between electrons.
In this work, however, we do not specify the form factor, since we would like to see whether and how well a boson exchange process can capture the momentum dependence of the four-fermion coupling constant.

For the superconducting boson we may try a simple ansatz for the propagator similar to \cref{equ:propagator_boson_af},
\begin{align} \label{equ:propagator_boson_cooper}
  \left(G^{(\varphi)}_k(Q)\right)^{-1}&= m_{\varphi,k}^2 + A_{\varphi ,k} [Q]^2\, ,
\end{align}
and take a momentum independent Yukawa coupling $h_{\varphi} = 1$.
The boson exchange contribution to the four-fermion interaction is computed analogously to \cref{main result}.
The solution to the field equation of $\varphi$ reads as
\begin{align} \label{equ:boson_cooper_eom}
  \varphi(Q) &= G_k^{(\varphi)}(Q) \sum_{P} \mathop{\Delta}\left(\vect{p} - \frac{\vect{q}}{2}\right) \psi_\alpha(P) \varepsilon_{\alpha \beta} \psi_\beta(Q-P)\,,
\end{align}
with $\vect{q}$ the (space-like) momentum component of $Q$.
Insertion of this solution into the effective action induces the boson exchange interaction
\begin{align} \label{equ:action_term_cooper_exchange}
  \Gamma_k^{\varphi} &= \sum_{Q, P_i} G_k^{(\varphi)}(Q) \mathop{\Delta}\left(\vect{p}_1 - \frac{\vect{q}}{2}\right) \mathop{\Delta}\left(\vect{p}_2 - \frac{\vect{q}}{2}\right) \nonumber \\
  &\phantom{=\;\;} \times \psi^{*}_\alpha(P_1) \varepsilon_{\alpha \beta} \psi^{*}_{\beta}(Q-P_1) \psi^{}_{\gamma}(P_2) \varepsilon_{\gamma \delta} \psi^{}_{\delta}(Q-P_2) \,.
\end{align}
This can be brought into the form~\eqref{equ:action_int} by use of the identity
\begin{align} \label{equ:epsilon_identity}
  \varepsilon_{\alpha \beta} \varepsilon_{\gamma \delta} &= \delta_{\alpha \gamma} \delta_{\beta \delta} - \delta_{\alpha \delta} \delta_{\beta \gamma} \,,
\end{align}
such that the boson exchange contribution to $U$ reads as
\begin{align} \label{equ:boson_exchange_cooper}
  &\phantom{=\;\;} \lambda_k^{(\varphi)}(P_1,P_2,P_3,P_4) \nonumber \\
  &= - 4 G_k^{(\varphi)}(P_1+P_3) \mathop{\Delta}\left(\frac{\vect{p}_3 - \vect{p}_1}{2}\right) \mathop{\Delta}\left(\frac{\vect{p}_4 - \vect{p}_2}{2}\right)
\end{align}

For the CPH channel with $(P,P,-P,-P)$, $P=(0,\vect{p})$, we have
\begin{align}
\label{eq: four fermion vertex of CPH in bosonisation}
  \lambda^{(\varphi)}_{\text{CPH}}(P) &= - 4 G_k^{(\varphi)}(0) \Delta^2(-\vect{p}) \,.
\end{align}
In the DPH channel with $(P,-P,-P,P)$ we find $U^{(\varphi)}_{\text{DPH}}(P) = U^{(\varphi)}_{\text{CPH}}(P)$, since $\mathop{\Delta}(\vect{p}) = \mathop{\Delta}(-\vect{p})$.
The PP channel with $(P,P,P,P)$ yields a contribution 
\begin{align}
  \lambda^{(\varphi)}_{\text{PP}}(P) = - 4 G_k^{(\varphi)}(2 P) \Delta^2(0)\,,
\end{align}
which vanishes for the $d_{x^2-y^2}$ and $d_{x y}$ waves.
For the $s$ waves, $\Delta(0) = 1$.

\subsection{Towards antiferromagnetism and superconductivity}
\label{Sect: Towards antiferromagnetism and superconductivity}
This work is limited to an investigation of the validity of partial bosonisation.  It seems useful, nevertheless, to sketch our proposal for continuing the flow into the ordered phase, including parts of the parameter space for which superconductivity is expected.
This proposal follows for the boson-exchange part Refs.\,\cite{Friederich:2010hr,Friederich:2010zn}, while we suggest to keep, in addition, a momentum dependent $\lambda_k^{(\psi)}$ instead of the constant used in Refs.\,\cite{Friederich:2010hr,Friederich:2010zn}.
For this purpose one will need an ansatz for $\Gamma_k^\text{boson}$ in \cref{equ:bosonised action_effective}.

The bosonic part $\Gamma_k^\text{boson}$ includes pure bosonic interactions involving the kinetic terms of $a^i$ and $d$.
We can parametrise it as
\begin{align}
\Gamma_k^\text{boson}=\sum_X V_k\fn{\alpha, \delta}+(\text{derivative terms})\,,
\end{align}
where the effective potential $V_k$ in coordinate space-time $X=(t,{\bvec x})$ is given by
\begin{align}
\label{equ:effective_boson_potential}
V_k\fn{\alpha,\delta}=m^2_{a,k} \,\alpha + m^2_{\varphi,k} \delta + \frac{\lambda_{a,k}}{2}\alpha^2+\frac{\lambda_{\varphi,k}}{2}\delta^2+\cdots\,,
\end{align}
with $\alpha\fn{X}=a^i\fn{X}a^i\fn{X}/2$ and $\delta(X) = \varphi^{*}(X) \varphi^{}(X)$.
This effective potential exhibits SU(2) (spin-rotation) symmetry.

Spontaneous symmetry breaking in the antiferromagnetic channel occurs if $m_a^2$ is negative for $k=0$, or more precisely for $k^{-1}$ of the order of the characteristic size of the probe.
In this case, the mass parameter $m^2_{a,k}$ varies from a positive value to a negative one through zero.
For $m_a^2(k = k_t) = 0$ the four-fermion interaction strength $U_{k_t}$ diverges.
This divergence, seen in many FRG investigations in the purely fermionic setting, indicates the onset of local symmetry breaking, typically with domain size $k_t^{-1}$.
For a negative value of the mass parameter $m_{a,k}^2<0$, the effective potential has a minimum at non-zero field value.
If this persists for $k\to0$, this is a signal of antiferromagnetic ordering $\langle a^i \rangle=\langle \psi^{\dag} \sigma^i\psi \rangle\neq 0$, as a consequence of SU(2) spin-rotation symmetry breaking.

For the derivative terms involving the kinetic terms of the bosonic fields $a^i$ and $\varphi$, we could make several choices.
A particularly simple choice amounts to assuming a separate and quadratic dependence on frequencies and momenta:
\begin{align} \label{equ:boson_kinetic_term}
  G_k^{-1}(Q) &= Z_k \omega_Q^2 + A_k [\vect{q}]^2+m_k\,.
\end{align}
Here $[\vect{q}]^2 = \vect{q}^2$ if $\vect{q} \in [-\pi,\pi]^2$, and is periodically continued otherwise.
$Z_k$ and $A_k$ are field renormalisation factors.
At the initial scale $k=\Lambda$, only the fermionic dynamics is relevant, so that we can infer $Z_\Lambda=A_\Lambda=0$ as initial conditions.
Positive finite values of $Z_k$ and $A_k$ would be generated by fluctuation effects in low-energy regimes, and then the bosonic field behaves as a dynamical collective mode.

Based on these general properties, we propose for the future a method for which explicit bosonic fields account for leading boson-exchange channels in the four-fermion interaction.
The part $\lambda^{(b)}_k$ corresponding to these leading channels can be shifted to the bosonic sector at every scale $k$ by dynamical bosonisation~\cite{Gies:2001nw, Gies:2002hq, Pawlowski:2005xe,
  Floerchinger:2009uf, Braun:2014ata, Fu:2019hdw}.
The residual part $U^{(0)}_k + \Delta\lambda^{(\psi)}_k(P_1, P_2, P_3,P_4)$ should be kept in the truncation, with bosonic contributions adding to the fermionic fluctuation effects.
The total four-fermion interaction $U$ can then be reconstructed as
\begin{align}
&U_k\fn{P_1,P_2,P_3,P_4}\nn
&\phantom{U} = \lambda_k^{(b)}(P_1,P_2,P_3,P_4) + \Delta\lambda_k^{(\psi)}(P_1,P_2,P_3,P_4) + U_k^{(0)} \;.
\label{Eq: bosonisation for four fermion vertex}
\end{align}
The advantage of this method will be an enhanced resolution of the leading channels by taking into account the effective boson-potential~\eqref{equ:effective_boson_potential} terms beyond the quadratic order.
This procedure allows to explore the regions with local or global spontaneous symmetry breaking without being stopped by the divergence of the four-fermion interaction in a purely fermionic treatment.

\subsection{Bosonisation in different momentum channels}\label{sec:propagator_fit_results}
A given choice of bosonisation is reflected in all different momentum channels.
For example, in our approximation the antiferromagnetic boson exchange leads in the particle-particle channel with $(P,P,P,P)$ to a momentum-independent constant.
Subtracting this constant in the PP channel will not absorb much of the four-fermion interaction.
This is demonstrated in \cref{fig:ff_coupling_IR_PP}.
Indeed, $U_{\text{PP}}(P) = U_k(P,P,P,P)$ does not exhibit the pronounced antiferromagnetic peaks visible in \cref{fig:ff_coupling_IR_main_figure,fig:ff_coupling_IR_PHD}.
It still shows some structure, which one may attribute to the exchange of other collective bosons.

In general, one may combine bosons in two or even more channels in order to account for the features in different momentum channels.
The total four-fermion vertex $U_k$ is then composed of different boson-exchange combinations $\lambda_k^{(i)}$, plus a residual contribution $\lambda_k^{(\psi)}$.
We observe that some boson-exchange channels give negative contributions to $U$, as seen for the particle-pair bosons in \cref{equ:boson_exchange_cooper} if the product of the form factors $\Delta$ is positive.
Another negative contribution is due to density waves.
Generalising \cref{equ:bosonised action_bare} by replacing $m^2$ with $\left(G^{(\rho)}_{k}(Q)\right)^{-1}$ one finds a contribution to the four-fermion interaction
\begin{align} \label{equ:boson_exchange_density}
  \lambda_k^{(\rho)}(P_1,P_2,P_3,P_4) &= - G_k^{(\rho)}(P_2-P_1) \,,
\end{align}
where we have again assumed $h_{\rho,k} = 1$.
The combination of attractive and repulsive contributions gives flexibility to the bosonisation procedure, but makes it also sometimes difficult to disentangle different effects.

As an example, we may look at the PP channel with momenta $(P,P,P,P)$. We observe in \cref{fig:ff_coupling_IR_PP} a momentum structure very different from \cref{fig:ff_coupling_IR_main_figure,fig:ff_coupling_IR_PHD}. 
If we want to associate this to boson-exchange peaks, we would need a negative contribution, similar to the exchange of density wave bosons or pairing bosons. 
With negative peaks at multiples of $\pi$ we actually find a good fit for 
\begin{align}
U_k(P,P,P,P) = - 4 (m_k^2+A_k[2{\bvec p}^2])^{-1} + \lambda^{(0)}_\text{PP}\,.
\end{align}
The factor $4$ is somewhat arbitrary at this stage. It would correspond to the exchange of an $s$-wave or extended $s$-wave pairing boson according to \cref{equ:boson_exchange_cooper}, with the boson propagator \eqref{equ:propagator_boson_cooper}.

To check the consistency of such a hypothesis, one has to look at the corresponding contribution in the CPH and DPH channels, for which the momentum dependence is given by the squared form factor in \cref{eq: four fermion vertex of CPH in bosonisation}.
For the $s$-wave pairing boson with $\Delta^2\fn{-{\bvec p}}=1$ we obtain in these channels a constant contributing to $U_k^{(0)}$,
\begin{align}
\Delta U_k^{(0)}=-\frac{4}{m_\varphi^2}\approx -1.66 \,,
\end{align}
where we use the fit value from the PP channel $m_\varphi^2=2.41$. This would imply for the constant part not indicated by the exchange of the antiferromagnetic or pairing boson a value $U_\text{CPH}^{(0)}\approx U_\text{DPH}^{(0)} \approx 1.08+1.66  = 2.74$.
This agrees reasonably well with the constant $U_\text{PP}^\text{(0)} \approx 2.42$ extracted from the fit in the PP channel.
We conclude that the simultaneous exchange of an antiferromagnetic and an $s$-wave pairing boson accounts for a reasonable fit of most of the momentum structures observed in all three investigated channels.
Of course, in order to confirm such a hypothesis, other momentum channels should be explored as well.

If we replace the $s$ wave by an extended $s$ wave we would predict in the CPH and DPH channels an additional momentum dependence mediated by the exchange of the pairing boson,
\begin{align}
\lambda^{(\varphi)}_\text{CPH} =\lambda^{(\varphi)}_\text{DPH} =-\frac{1}{m_\varphi^2} \left( \cos\fn{p_x}+\cos\fn{p_y}\right)^2\,.
\end{align}
This should be visible in $\Delta \lambda_k^{(\psi)}\fn{\bvec p}$ in \cref{fig:ff_coupling_IR_main_figure,fig:ff_coupling_IR_PHD} (lower right corner).
The negative contribution is maximal for $p_x=p_y=0,\,\pm\pi$.
This is not the observed structure, and we thus conclude that an extended $s$-wave boson does not play an important role. Our discussion exemplifies that the choice of appropriate bosons proceeds often by hypothesis and verification/falsification, rather than by a systematic treatment.

If we follow the hypothesis of a boson exchange contribution from both an antiferromagnetic and $s$-wave pairing bosons we arrive in the three channels at a description of the four-fermion interaction by
\begin{align}
&U_{\text{CPH}}(\vect{p}) = 2\, G_k^{(a)}( 2 \vect{p} + \vect{\pi} ) + c_k^{(a)} -\frac{4}{m_\varphi^2}  \nn
&\phantom{U_{\text{CPH}}(\vect{p}) =}+ U^{(0)}_{\text{CPH}} + \Delta\lambda^{(\psi)}_\text{CPH}\fn{\vect{p}}\,, \nn[1em]
&U_{\text{DPH}}(\vect{p}) = G_k^{(a)}( 2 \vect{p} + \vect{\pi}) +2c_k^{(a)}-\frac{4}{m_\varphi^2} \nn
&\phantom{U_{\text{DPH}}(\vect{p}) = } + U^{(0)}_{\text{DPH}} + \Delta\lambda^{(\psi)}_\text{DPH}\fn{\vect{p}}\,,\nn[1em]
&U_{\text{PP}}(\vect{p}) = -4 \, G_k^{(\varphi)}( 2 \vect{p}) +3c_k^{(a)} + U^{(0)}_{\text{PP}} + \Delta\lambda^{(\psi)}_\text{PP}(\vect{p})\,,
\label{eq: simplified bosonisation}
\end{align}
with
\begin{align}
c_k^{(a)} = G^{(a)}_k(\vect{\pi}) =  \frac{1}{m_{a,k}^2+2\pi^2 A_{a,k}}\,.
\end{align}
Here $G_k^{(a)}$ and $G_k^{(\varphi)}$ are both given by \cref{equ:propagator_boson_cooper}, but with different parameters $m_k^2$ and $A_k$.

For a fit to the four-fermion interaction at the scale $k_\text{IR}$ we combine all constants to a common constant $\lambda_k^{(0)}$ and neglect $\Delta\lambda_k^{(\psi)}$.
In each channel we therefore perform a fit with three parameters $m_k^2$, $A_k$ and $\lambda_k^{(0)}$.
We obtain the following fit parameters:
\begin{align} \label{equ:fit_params}
   \left( m_k^2, A_k, \lambda_k^{(0)} \right)_{\text{CPH}} &\approx ( 0.32, 0.53, 1.17 ) \,, \nonumber \\[1ex]
   \left( m_k^2, A_k, \lambda_k^{(0)} \right)_{\text{DPH}} &\approx ( 0.41, 0.69, 1.22 )\,, \nonumber \\[1ex]
   \left( m_k^2, A_k, \lambda_k^{(0)} \right)_{\text{PP}} &\approx ( 2.41, 2.29, 2.68 )\,.
\end{align}
The parameters $m_k^2$ and $A_k$ should be similar for the CPH and DPH channels, since both involve the same propagator $G_k^{(a)}$ for the antiferromagnetic boson.
The parameters in the PP channel are different. In particular, we observe $m_{a,k}^2\ll m_{\varphi,k}^2$, indicating that the antiferromagnetic channel dominates.

With the ``measured values" $m_{i,k}^2$, $A_{i,k}$, $\lambda_{i,k}^{(0)}$ we can compute the residual constant part $U_k^{(0)}$ in the four-fermion interaction after subtraction of the constant boson exchange contributions.
From \cref{eq: simplified bosonisation} one finds
\begin{align}
  U^{(0)}_{\text{CPH}} \approx 2.76\;, \quad U^{(0)}_{\text{DPH}} \approx 2.69, \quad U^{(0)}_{\text{PP}} &\approx 2.42\;.
\end{align}
The values are actually rather similar and close to three, indicating that at least for the considered channels the fluctuation contribution is dominated by boson exchange.

As can be seen in the upper right in \cref{fig:ff_coupling_IR_main_figure,fig:ff_coupling_IR_PHD}, our fit works well in the particle-hole channels, where it captures the strong nesting peaks signifying the neighbourhood of the onset of antiferromagnetic ordering quite nicely.
In addition, the particle-particle channel exhibits a repulsive behaviour, which can also be fitted by a $s$-wave pairing boson exchange, as shown in \cref{fig:ff_coupling_IR_PP}.
The residual four-fermion interactions shown in the bottom of these figures are in all cases significantly smaller than the main features absorbed into the boson propagators.
We also note that some of the differences could be artefacts stemming from the $N$-patching approximation we used to feed back the momentum dependence of the four-fermion coupling, or other truncation effects.
In addition, if other effective boson-exchange processes like, e.g., a charge density wave played a role at half-filling, they should be visible in at least one of the channels we investigated.
Since we observe no large additional contribution of this type, we can conclude that these processes as expected indeed do not play an important role in the transition to antiferromagnetic order, {\it a posteriori} justifying our ansatz for bosonising the four-fermion coupling in \cref{Sect: Towards antiferromagnetism and superconductivity}.

In summary, we find that the leading instabilities in the four-fermion interaction which signal the transition to antiferromagnetic order can be described as exchanges of an antiferromagnetic spin wave.
While some substructures in the fermionic two-particle vertex are not absorbed by our simple bosonisation ansatz, these are small compared to the nesting peaks in the particle-hole channels, and might at least partially be truncation artefacts.
We thus conclude that it is indeed possible to bosonise the Hubbard model close to half filling.

\subsection{Build-up of bosonic exchange channels} \label{sec:G_of_k}
\begin{figure}[tbp]
  \centering
  \includegraphics[width=\linewidth]{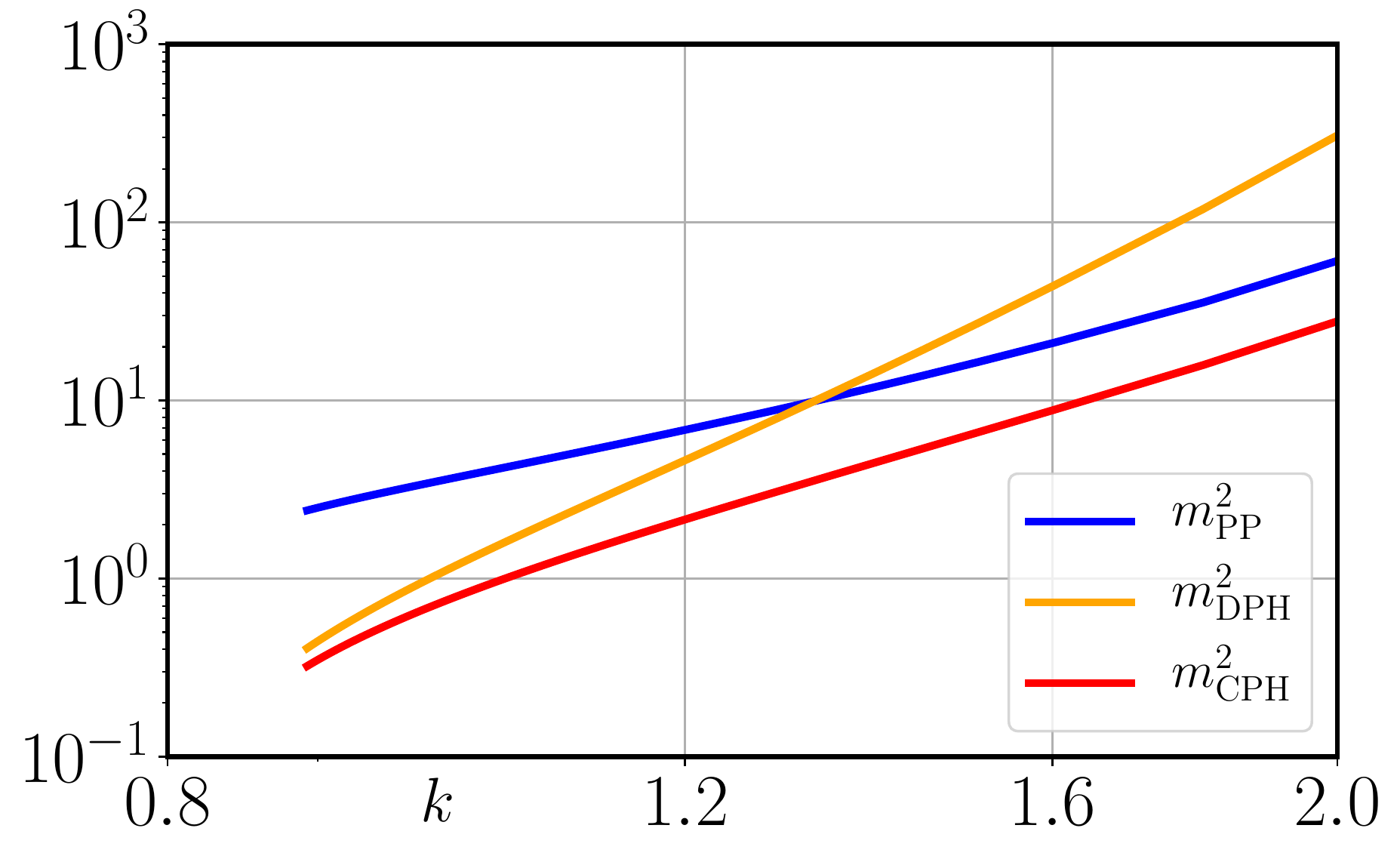}
  
  \includegraphics[width=\linewidth]{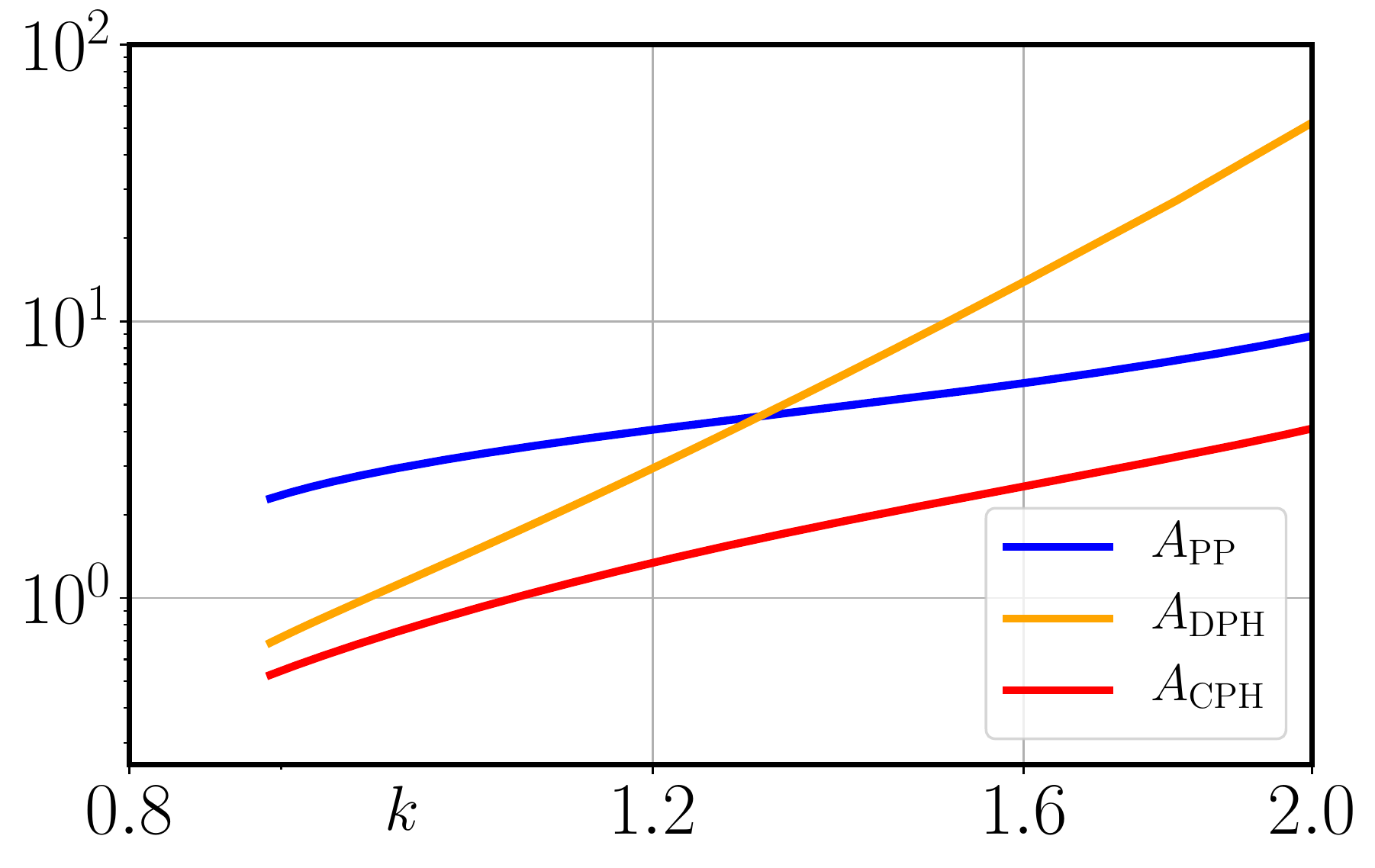}
  
  \caption[Propagator fit parameters as a function of k]{
  Fit parameters $m_k^2$ and $A_k$ for the boson-exchange propagators presented in \cref{sec:propagator_fit_results} at different renormalisation scales $k$.
  In all cases the values decrease exponentially towards the IR, signalling a build-up of nesting peaks, or more generally a non-trivial momentum dependence of the four-fermion coupling which can be described as an effective boson exchange.
 The parameters for the two particle-hole channels tend towards each other, consistent with the exchange of an antiferromagnetic boson in the two channels.
   }
  \label{fig:fits_of_k} 
\end{figure}
In order to investigate the build-up of antiferromagnetic ordering, we repeat the fits presented in \cref{sec:propagator_fit_results} at different scales $k$, starting from the initial UV cutoff $\Lambda$ (here $\Lambda = 6 t$).
We present the resulting scale dependence of the fit parameters $m_k^2$ and $A_k$ in \cref{fig:fits_of_k}.
As expected, close to the microscopic UV scale the values for $m_k^2$ and $A_k$ are very large, corresponding to a negligible boson-exchange contribution.
The fit parameters become smaller towards the IR, however, signalling a build-up of nesting peaks in the particle-hole channels, and a non-trivial momentum dependence in the particle-particle channel.
Close to the transition to antiferromagnetic ordering, where our flow stops numerically, the fit parameters for the particle-hole channels seem to flow towards the same values.
We interpret this as a signal that the parts of the four-fermion interaction which signal the transition to antiferromagnetic order can indeed be bosonised by a single antiferromagnetic exchange boson $a$.

We also observe that for large $k$ the effects of the exchange of antiferromagnetic and $s$-wave pairing bosons are of similar size. Only close to the onset of local antiferromagnetic order a clear dominance of the antiferromagnetic channel sets in.

\section{Conclusions} \label{sec:conclusions}
In this paper, we investigated the momentum dependence of the four-fermion coupling of the two-dimensional Hubbard model on a square lattice close to the transition to antiferromagnetic order.
We found that the nesting singularities in the particle-hole channels driving the transition can be well described by the exchange of an antiferromagnetic spin wave.
This already works for a particularly simple parametrisation of the exchange-boson propagator.
At half-filling the residual four-fermion interaction is small compared to the part that can be described by boson exchange.

This suggests a split of the momentum-dependent four-fermion interaction $U_k$ into a boson-exchange part $\lambda_k^{(b)}$, and a residual four-fermion interaction $\lambda_k^{(\psi)}$.
This can be done at every scale $k$ of the renormalisation flow, i.e.,
\begin{align}
  U_k(P_i) = \lambda^{(b)}_k(P_i) + \lambda_k^{(\psi)}(P_i) \,.
\end{align}
The boson-exchange part can be described by explicit bosonic fields for collective excitations using the method of flowing bosonisation.
Keeping a momentum-dependent residual interaction $\lambda_k^{(\psi)}$ is important in order to diminish the effects of a possible bias in the choice of the bosonisation, maintain an accurate momentum dependence, and detect possible other channels for which $U_k$ can get large, and which are not yet included in $\lambda^{(b)}_k$.

There is no need to restrict this method of partial bosonisation to a single bosonic field. For two or more bosonic fields, effects of competing channels, competing order, or simultaneous order for several order parameters, can be investigated.
For the present investigation of three-momentum channels we find a good description of the momentum-dependent four-fermion vertex by the exchange of an antiferromagnetic and an $s$-wave pairing boson.
The simultaneous flow of interactions for fermionic and bosonic degrees of freedom enhances the algebraic and numerical complexity.
However, the bonus of access to the ordered phase and the increased resolution for the leading channels makes it worth it.

\subsection*{Acknowledgements}
We thank Michael Scherer for discussions. This work
is supported by EMMI, and is part of and supported by the DFG Collaborative Research Centre SFB 1225 (ISOQUANT) as well as by the
DFG under Germany's Excellence Strategy Grant No. EXC - 2181/1 - 390900948 (the
Heidelberg Excellence Cluster STRUCTURES). M.\,Y. is supported by the Alexander
von Humboldt Foundation.

\appendix
\begin{widetext}
\section{Flow equation} \label{sec:flow_equations}
In this appendix, we show the explicit form of the beta function of the four-fermion coupling
$U_k$ occurring in the effective action \eqref{equ:action_eff}.
A key quantity in the flow equation is the two-point function $\Gamma_k^{(2)}$.
In the Hubbard model, we compute the second-order functional derivative with respect to $\psi$ and $\bar\psi$.
Its result is shown in e.g. \cite{Kopietz:2010zz}.
In order to obtain the flow of $U_k$, we define the spin
projector
\begin{align} \label{equ:four-fermi_proj}
  \mathds{P}_{s_1 s_2 s_3 s_4} &= \frac{1}{2} \left( \delta_{s_1 s_2} \delta_{s_3 s_4} - \delta_{s_1 s_3} \delta_{s_2 s_4} \right)\,.
\end{align}
The flow equation of the four-fermion vertex $U_k$ can then be obtained by acting with this projector on the flow equation of the four-point function.
In summary, we arrive at
\begin{align} \label{equ:four-fermi_vertex}
  \partial_k U_k(P_1,P_2,P_3,P_4)
 & = \text{Tr}\left(
   \mathds{P}\frac{\delta^4 \p_k \Gamma_k }{ \delta\psi(-P_4)\delta\psi^{\dag}(P_3)\delta\psi (-P_2)\delta\psi^{\dag}(P_1) }\right) \nonumber \\[1ex]
   &= \left( \mathcal{T}_{\text{PP}} + \mathcal{T}_{\text{DPH}} + \mathcal{T}_{\text{CPH}} \right) \delta(P_1-P_2+P_3-P_4) \;,
\end{align}
where we have used the flow equation~\eqref{equ:wetterich_equation} in the second line. 
The spin-projected flow diagrams $\mathcal{T}_{i}$ read as~\cite{Salmhofer:2001tr,PhysRevB.64.155101}
\begin{align} 
  \mathcal{T}_{\text{PP}}(P_1,P_2,P_3,P_4) &= - \sum_Q L(-Q, Q + P_1 + P_3) U_k(P_1,-Q,P_3,Q+P_1+P_3) U_k(-Q,P_2,Q+P_1+P_3,P_4) \,,\nonumber \\[0.5em]
  \mathcal{T}_{\text{DPH}}(P_1,P_2,P_3,P_4) &= - \sum_Q L(Q, Q+P_1-P_2) \Big( - 2 U_k(P_1,P_2,Q,Q+P_1-P_2) U_k(Q+P_1-P_2,Q,P_3,P_4) \nonumber \\
  &\phantom{=\;}\hspace{4.8cm} + U_k(P_1,Q+P_1-P_2,Q,P_2) U_k(Q+P_1-P_2,Q,P_3,P_4) \nonumber \\
  &\phantom{=\;}\hspace{4.8cm} + U_k(P_1,P_2,Q,Q+P_1-P_2) U_k(P_3,Q,Q+P_1-P_2,P_4) \Big)\,, \nonumber \\[0.5em]
  \mathcal{T}_{\text{CPH}}(P_1,P_2,P_3,P_4) &= - \sum_Q L(Q, Q-P_1+P_4) U_k(P_1,Q,Q-P_1+P_4,P_4) U_k(Q,P_2,P_3,Q-P_1+P_4) \;.
  \label{equ:flow_equations}
\end{align}
The abbreviations ``PP", ``DPH'', and ``CPH" stand for particle-particle, direct particle-hole and crossed particle-hole channels, respectively, 
and we introduce the threshold function
\begin{align}
\label{App; eq: threshold function}
L(P,Q) \coloneqq \partial_k \left( G_k^\text{f}(P) \,  G_k^\text{f}(Q) \right)\,,
\end{align}
with the full fermion propagator
\begin{align} \label{equ:prop_fermion}
  G^{\text{f}}_k(Q) &= \frac{1}{\imaginaryi\, \omega_Q + \xi_{\vect{q}} + R^{\text{f}}_k(Q)} \;.
\end{align}

\section{Matsubara-resummed propagator loops} \label{sec:matsubara_sums}
The flow generators~\eqref{equ:flow_equations} involve Matsubara sums.
In this work, as discussed in \cref{Sec: solving pure fermionic equation}, we approximate the four-fermion vertex as frequency-independent and can therefore perform the Matsubara sums analytically.
In particular, with $\omega_P = 2 \pi T_k ( n_P + \frac{1}{2}) = \pm \omega_Q$, where $ n_{P} \in \mathbb{Z}$ and the effective temperature cutoff \eqref{equ:regulator_fermion}, we obtain
\begin{align}
 &T \sum_{n \in \mathbb{Z}} L(P,Q) 
  = \partial_k \left( T \sum_{n \in \mathbb{Z}} G_k^{\text{f}}(P) G_k^{\text{f}}(Q) \right) 
  = T \partial_k \left( - \cfrac{1}{2 T_k} \cfrac{\tanh\left(\cfrac{\xi_{\vect{p}}}{2 T_k}\right) - \tanh\left(\cfrac{\xi_{\vect{q}}}{2 T_k}\right)}{ \xi_{\vect{p}} - \xi_{\vect{q}} } \right)\,, \nonumber \\[2ex]
 &T \sum_{n \in \mathbb{Z}} L(P,-Q) 
 = \partial_k \left( T \sum_{n \in \mathbb{Z}} G_k^{\text{f}}(P) G_k^{\text{f}}(-Q) \right) 
 = T \partial_k \left( \cfrac{1}{2 T_k} \cfrac{\tanh\left(\cfrac{\xi_{\vect{p}}}{2 T_k}\right) + \tanh\left(\cfrac{\xi_{\vect{q}}}{2 T_k}\right)}{ \xi_{\vect{p}} + \xi_{\vect{q}} } \right) \,.
\label{eq: Matsubara summation}
\end{align}
We specified our choice of the fermion regulator and the precise form of the effective temperature $T_k$ in \cref{sec:fermion_regulator}.
\end{widetext}

\section{Computation cycle}
We briefly summarise the computation cycle used in this paper.
\begin{enumerate}
\item Derive beta functions by taking functional derivatives of exact flow equation.
\item Solve the beta function of the four-fermion coupling, $\p_k U_k(P_1,P_2,P_3,P_4)=\mathcal{T}_{\text{PP}} + \mathcal{T}_{\text{DPH}} + \mathcal{T}_{\text{CPH}}$ for three different momentum configurations (DPH, CPH, and PP) by using $N$-patching method. (The top left-hand side figures in \cref{fig:ff_coupling_IR_main_figure,fig:ff_coupling_IR_PHD,fig:ff_coupling_IR_PP}.)
\item We fit to the obtained $U_k(P_i)=U_k(P_1,P_2,P_3,P_4)$ in step~2 by the bosonic propagator,
\begin{align}
&\lambda_k^{(b)}(P_i) \nn
&\quad=\frac{n_1}{m_k^2+A_k[2{\vect p}-{\vect\pi}]^2}\
+\frac{n_2}{m_k^2+2\pi^2A_k}+U_k^{(0)}\,,
\end{align}
and find $m_k^2$, $A_k$ and $U_k^{(0)}$. (The top-right hand side figures in \cref{fig:ff_coupling_IR_main_figure,fig:ff_coupling_IR_PHD,fig:ff_coupling_IR_PP}.)
\item We evaluate the difference between $U_k(P_i)$ and $\lambda_k^{(b)}$,
\al{
\Delta\lambda_k^{(\psi)}(P_i) = U_k(P_i)-\lambda_k^{(b)}(P_i)\,.
}
(The bottom hand side figures in \cref{fig:ff_coupling_IR_main_figure,fig:ff_coupling_IR_PHD,fig:ff_coupling_IR_PP}.)
\end{enumerate}

\bibliography{hubbard}

\end{document}